\documentclass[letterpaper]{article} % DO NOT CHANGE THIS
\usepackage{aaai24}  % DO NOT CHANGE THIS
\usepackage{times}  % DO NOT CHANGE THIS
\usepackage{helvet}  % DO NOT CHANGE THIS
\usepackage{courier}  % DO NOT CHANGE THIS
\usepackage[hyphens]{url}  % DO NOT CHANGE THIS
\usepackage{graphicx} % DO NOT CHANGE THIS
\usepackage{natbib}  % DO NOT CHANGE THIS AND DO NOT ADD ANY OPTIONS TO IT
\usepackage{caption} % DO NOT CHANGE THIS AND DO NOT ADD ANY OPTIONS TO IT
\usepackage{algorithm}
\usepackage{algorithmic}

\usepackage{booktabs}
\usepackage{multirow}
\usepackage{colortbl}
\usepackage{nameref}
\usepackage{url}

\usepackage{newfloat}
\usepackage{listings}
\DeclareCaptionStyle{ruled}{labelfont=normalfont,labelsep=colon,strut=off} % DO NOT CHANGE THIS
\floatstyle{ruled}
\newfloat{listing}{tb}{lst}{}
\floatname{listing}{Listing}
\nocopyright
\title{Co-designing an AI Impact Assessment Report Template \\ with AI Practitioners and AI Compliance Experts}

\author{
    Edyta Bogucka\textsuperscript{\rm 1}, Marios Constantinides\textsuperscript{\rm 1}, Sanja \v{S}\'{c}epanovi\'{c}\textsuperscript{\rm 1},
    Daniele Quercia\textsuperscript{\rm 1,2}\\
}
\affiliations {
    \textsuperscript{\rm 1}Nokia Bell Labs, Cambridge, UK\\
    \textsuperscript{\rm 2}Kings College London, London UK\\
    edyta.bogucka@nokia-bell-labs.com.com, marios.constantinides@nokia-bell-labs.com, sanja.scepanovic@nokia-bell-labs.com,
    daniele.quercia@nokia-bell-labs.com
}

\begin{document}

\maketitle

\begin{abstract}
In the evolving landscape of AI regulation, it is crucial for companies to conduct impact assessments and document their compliance through comprehensive reports. However, current reports lack grounding in regulations and often focus on specific aspects like privacy in relation to AI systems, without addressing the real-world uses of these systems. Moreover, there is no systematic effort to design and evaluate these reports with both AI practitioners and AI compliance experts. To address this gap, we conducted an iterative co-design process with 14 AI practitioners and 6 AI compliance experts and proposed a template for impact assessment reports grounded in the EU AI Act, NIST's AI Risk Management Framework, and ISO 42001 AI Management System. We evaluated the template by producing an impact assessment report for an AI-based meeting companion at a major tech company. A user study with 8 AI practitioners from the same company and 5 AI compliance experts from industry and academia revealed that our template effectively provides necessary information for impact assessments and documents the broad impacts of AI systems. Participants envisioned using the template not only at the pre-deployment stage for compliance but also as a tool to guide the design stage of AI uses.
\end{abstract}

\makeatletter
\let\@makefntext@orig\@makefntext
\renewcommand\@makefntext[1]{#1}
\makeatother

\begingroup
\renewcommand\thefootnote{}
\footnotetext{Project website: \textbf{https://social-dynamics.net/impact-assessment}}
\endgroup

\makeatletter
\let\@makefntext\@makefntext@orig
\makeatother

\section{Introduction}
\label{sec:intro}

The potential of AI to bring about significant societal shifts requires a careful examination of the associated risks and benefits \cite{unesco2023ethicalImpactAssessment}. This has prompted scholars to investigate established impact assessments processes in fields such as environmental protection \cite{selbst2021algorithmicImpact, metcalf2021constructingImpacts} and human rights \cite{hria_2023, unesco2023ethicalImpactAssessment}, as well as to suggest algorithmic impact assessments (AIAs) as initial self-regulatory approaches for recognizing and mitigating algorithmic harms \cite{metcalf2021constructingImpacts, pwc2021impactAssessment}. These assessments, guided by ethical principles such as responsibility and fairness, were recommended to be conducted at different stages, such as system's design, pre-launch, and post-launch, and to be publicly shared as impact statements \cite{diakopoulos2016socialImpactStatement}.

As it has become evident that AI system impacts involve not only the underlying algorithms \cite{Watkins2021, Barnett2022} but also their systemic impacts \cite{ehsan2022algorithmicImprint, shelby2023sociotechnicalHarms}, there has been a growing demand for comprehensive AI impact assessments (AIIAs) \cite{selbst2021algorithmicImpact}. In the absence of regulatory requirements and standardized frameworks, initially, AIIAs have been regarded as an extension of existing AI governance processes, such as assessments of privacy \cite{wright2012privacyImpactAssessment}, data protection \cite{fundamentalRightsList_2020}, and social consequences of AI systems, models and services \cite{selbst2021algorithmicImpact, raji2020accountabilityGap}. However, over time, AIIAs developed into an independent component of AI governance \cite{skoric2023assessmentCriteria}. By 2021, a total of 38 distinct impact assessment methods have been introduced \cite{stahl2023systematicReview} serving various goals such as prompting companies to proactively address social consequences \cite{sherman2023riskProfiles}, aligning system behavior with organization’s responsible AI principles \cite{raji2020accountabilityGap, fujitsu2022ethicalAssessment, microsoft2022raiImpactAssessmentTemplate}, increasing awareness of potential harms among development teams \cite{johnson2023classroomStudy}, and documenting decisions to facilitate learning and future governance development \cite{selbst2021algorithmicImpact}.

However, current AIIAs reports are marked by a lack of comprehensiveness and insufficient grounding in established frameworks such as the EU AI Act~\cite{EUACT2021}, the NIST's AI Risk Management Framework (AI RMF)~\cite{nist2023aiRisk}, and international standards such as the ISO 42001 AI Management System~\cite{iso2023ManagementSystem}, that provide complementary angles for impact assessment: regulatory compliance (EU AI Act), risk management (AI RMF), and organizational best practices (ISO 42001). The lack of alignment with regulations makes it harder for stakeholders to understand the broad societal, environmental, and economic impacts of AI uses, which are essential for risk assessments \cite{EUACT2021, Hupont2024, stahl2023systematicReview}. Instead, they must repeatedly gather information across disparate assessments of various AI components (e.g., models, systems, and algorithms) rather than focusing on specific uses. Developer teams, in particular, often encounter difficulties in initiating AI impact assessments \cite{bucinca2023harmGeneration} and require additional guidance throughout this process \cite{wang2023designing}. 

\clearpage
In response to these issues, our aim is to examine recent regulatory frameworks and existing AIIAs to devise together with AI practitioners and AI compliance experts a reporting template focusing on the intended use of the system. This template aims to provide the necessary information for conducting impact assessments based on regulations across various AI system uses and to be adaptable to different roles. In so doing, we made two contributions: 

\begin{enumerate}
    \item We designed a comprehensive template for an impact assessment report that is grounded in regulations, and did so by conducting two studies. First, we elicited an initial set of design requirements through literature review and semi-structured interviews with 2 AI compliance experts. Second, we co-designed the impact assessment report template with 14 AI practitioners and 6 AI compliance experts, grounded it in regulatory requirements of the EU AI Act, AI RMF, and ISO 42001.
    \item We populated the template with a real-world use case of an AI system (a companion app designed to improve the meeting experience) and evaluated it with 8 AI practitioners and 5 AI compliance experts from academia and industry. We compared the final report to a baseline report derived from the typical structure of existing impact assessment reports. Participants found that the final report provided more complete information for impact assessments and addressed all AI system components and impacts more broadly than the baseline. Both the final report and the baseline were rated as highly adaptable to different AI uses and adaptable to different roles.
\end{enumerate}

In light of these results, we discuss the implications of our work for designing impact assessment reports and conducting impact assessment. 
\section{Related Work}
We surveyed various lines of research that our work draws upon and grouped them into three main areas: \emph{(1)} eliciting requirements for designing a comprehensive template for an impact assessment report that is grounded in regulations, \emph{(2)} co-designing the template, and \emph{(3)} evaluating the template. 

\begin{figure*}[t!]
  \centering
  \includegraphics[width=\textwidth]{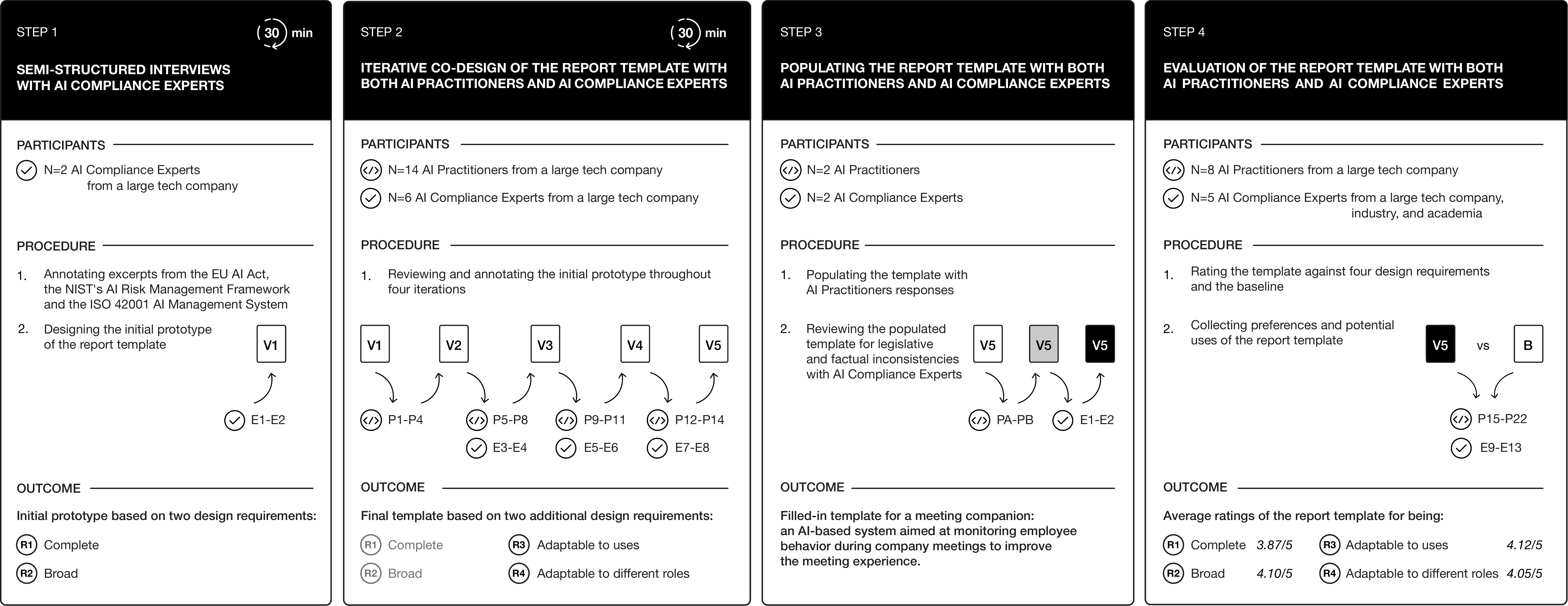}
  \caption{Overview of our four-step method for designing a comprehensive template for an impact assessment report grounded in the EU AI Act, NIST’s AI RMF, and ISO 42001. In the first step, we interviewed AI compliance experts to elicit two design requirements and design the initial prototype of the template (V1). In the second step, we engaged both AI practitioners and AI experts to iteratively elicit two additional design requirements and co-design four iterations of the template (V2-V5). In the third step, we populated the final version of the template with AI practitioners' responses. In the fourth step, we evaluated it against our four design requirements and the baseline template identified in the literature review of AIIAs \cite{stahl2023systematicReview}.}
  \label{fig:methodology}
\end{figure*}

\subsection{Requirements for Designing a Comprehensive Template That Is Grounded in Regulations}
AI impact assessments are defined as structured processes for understanding the implications of proposed AI systems \cite{stahl2023systematicReview}. They have been proposed as both law-agnostic self-regulating processes \cite{nist2023aiRisk, equalAI2023nist} and official processes in regulations \cite{EUACT2021, hria2023denmark, admTool2023canada} and organizational standards \cite{iso2023ManagementSystem}. For example, the NIST AI RMF, a voluntary guidance framework for organizations that design, develop, deploy, or use AI systems, suggests assessing the beneficial and harmful impacts of AI on individuals, groups, organizations, and society. The EU AI Act \cite{EUACT2021}, the first comprehensive AI regulation in the European Union, requires fundamental rights impact assessment for high-risk AI uses deployed by bodies governed by public law, private operators providing public services, and operators assessing creditworthiness and conducting risk assessments for life and health insurance. This impact assessment, conducted before deployment and updated as needed, should cover affected groups, risks of harm, human oversight, and risk management strategies. Similarly, the world's first ISO 42001 standard for AI management system \cite{iso2023ManagementSystem} requires organizations to assess the potential consequences of AI systems on individuals and societies, including human rights, legal positions, and life opportunities.

Despite ongoing efforts to standardize AIIAs \cite{iso2025aiSystemImpactAssessment}, a consensus on their content and reporting methods for stakeholders and the public is still lacking \cite{Watkins2021, sherman2023riskProfiles}. \citeauthor{stahl2023systematicReview} (\citeyear{stahl2023systematicReview}) identified and studied 38 AIIAs, showing that they vary widely in topics, focus, and formats, primarily covering human rights \cite{algorithmsHumanRights_2018, mantelero2022} and ethics \cite{gebru2021datasheets}. The focus of AIIAs varies, assessing the impacts of AI models, systems, or services, typically mentioning use cases only briefly in final reports. However, there is growing agreement that impact assessments should be conducted on specific uses \cite{EUACT2021, Hupont2024}. In the EU AI Act \cite{EUACT2021}, the specific use determines one of four risk categories (unacceptable, high, limited, minimal), each with its own legal requirements. Similarly, in the NIST framework \cite{nist2023aiRisk}, use-case profiles are required to describe the current state and the desired state of the system, allowing for risk management at different stages of the AI lifecycle.

\subsection{Collaborative Design of an Impact Assessment Report Template}
To achieve consensus on AIIAs, their report templates should adequately document AI uses, reflect the socio-technical nature of their impacts \cite{metcalf2021constructingImpacts, equalAI2023nist}, and accommodate different roles (e.g., developers, researchers, managers, compliance experts) to surface these impacts \cite{Moss2021assemblingAccountability, selbst2021algorithmicImpact, ALTAI_2020, raiCrafting}. However, existing AIIAs often lack publicly available documentation on their design processes \cite{johnson2023classroomStudy}, raising concerns about whether the current reports' templates meet the diverse needs of stakeholders. 

When available, the documentation typically only specifies the nature of the consultations with stakeholders, lacking depth about the types of stakeholders involved and how their input influenced the final design. For example, Microsoft's Responsible AI Impact Assessment template was developed following discussions with internal collaborators \cite{microsoft2022raiImpactAssessmentTemplate}, while the Canadian ADM template \cite{admTool2023canada} was revised through consultations with relevant stakeholders. In contrast, the Ada Lovelace Institute's Algorithmic Impact Assessment template in healthcare was developed through interviews with potential users from companies, researchers, and impact assessment experts \cite{ada_lovelace}. 

Researchers highlight the need for more inclusive, value-sensitive co-design processes~\cite{johnson2023classroomStudy, sadek2024guidelines} that explicitly incorporate compliance experts~\cite{Hupont2024, raiCrafting}. For example, OpenLoop's policy prototyping experiment involved developers, researchers, policymakers, and regulators collaborating to create a new policy and documentation of potential harms and mitigations for automated decision-making systems~\cite{Andrade2021}. Similarly, the creators of use case cards detailing AI systems' intended uses engaged in co-design workshops with Unified Modeling Language (UML) experts, and EU policy experts \cite{Hupont2024}.

\subsection{Evaluation of Impact Assessment Report Template}
Even less common than documenting template design processes is the public release of their evaluations \cite{johnson2023classroomStudy}, resulting in a lack of evidence for their effectiveness and contributing to the lack of consensus on what constitutes a comprehensive impact assessment \cite{schiff2020principles, stahl2023systematicReview}. User studies evaluating three impact assessment report templates, with students role-playing organizational roles, show that template design influences perceptions of AI risks and stakeholder responsibility for potential harm \cite{johnson2023classroomStudy}. The report's content can also affect stakeholders' perceptions and trust \cite{gaba2023visualDesign}, especially when it incorporates best practices from information visualization \cite{visWhatWorks2021}.

More broadly, AIIA evaluations have identified three key pitfalls of current templates. 
First, AIIAs are fragmented and templates fail to encompass the full range of impacts, including environmental concerns, democratic, safety, human agency, and economic factors \cite{stahl2023systematicReview}. Second, AIIAs are not tailored to multiple stakeholders, with templates featuring unfamiliar terms understandable only by those with technical knowledge or terms with multiple legal meanings \cite{schiff2020principles}. Third, AIIAs lack supporting questionnaires derived from legal frameworks \cite{skoric2023assessmentCriteria} to ensure templates meet two requirements: completeness of necessary information for risk assessment and broad overage of system components to identify socio-technical risks.

\smallskip
\noindent\textbf{Research Gaps.} Current impact assessment reports lack grounding in established regulations and often focus on specific aspects like human rights or data protection, without fully addressing all relevant compliance issues related to specific uses of AI systems. Moreover, there is no systematic effort to design and evaluate these reports with both AI practitioners and AI compliance experts.
\section{Methods}
\label{sec:report-template}
To design a comprehensive template for an impact assessment report that is grounded in regulations, we conducted a series of four studies (Figure \ref{fig:methodology}). These studies included literature review and interviews with 2 AI compliance experts, an iterative co-design process with 14 AI practitioners and 6 AI compliance experts, review of the populated template, and a user study to evaluate the populated template with 8 AI practitioners working in industry research and 5 AI compliance experts from industry and academia. 

All sessions were conducted via video conferencing over three months, and were recorded and transcribed. We ensured anonymity of data by excluding  personal identifiers, and maintained exclusive access to the data for the research team only. All sessions were approved by our organization.

\subsection{Eliciting Design Requirements From Literature Review and Semi-structured Interviews With AI Compliance Experts}
To elicit requirements for designing a comprehensive template for an impact assessment report that is grounded in regulations, we resorted to previous literature and did so in two steps. First, we reviewed prior academic literature investigating the use of risk assessments, impact reports, and AI documentation for compliance~\cite{skoric2023assessmentCriteria, selbst2021algorithmicImpact, hria_2023, nist2023aiRisk}. We also drew upon five existing AI impact assessment templates~\cite{microsoft2022raiImpactAssessmentTemplate, fujitsu2022ethicalAssessment, equalAI2023nist, ada_lovelace, admTool2023canada} and value-sensitive games and cards \cite{judgmentCall_2019, eccola2021, sadek2024guidelines}. Second, we identified the necessary minimum information to conduct impact assessment in line with three frameworks: the EU AI Act~\cite{EUACT2021}, the NIST's AI RMF~\cite{nist2023aiRisk}, and the ISO 42001 \cite{iso2023ManagementSystem}. We selected these frameworks because they have undergone a consensus-driven, transparent process with wide consultations involving various roles such as developers, researchers, managers, and compliance experts. Together, they provide three complementary angles for impact assessment: the EU AI Act offers regulatory compliance, setting a precedent for other AI regulations; the AI RMF provides practical insights for risk management in corporate environments; and ISO 42001 emphasizes organizational best practices. In this step, we started from reviewing the EU AI Act~\cite{EUACT2021} and identifying relevant excerpts that pertain to risk management (Article 9), data governance (Article 10), system monitoring (Articles 14, 15, 72), technical documentation (Articles 11, 12, 18, Annex IV), system transparency (Articles 13, 50), impact assessment (Article 27), and provider obligations (Articles 16, 17, 53, 55). We then reviewed the NIST's AI RMF~\cite{nist2023aiRisk} and its Algorithmic Impact Assessment template~\cite{equalAI2023nist}. From these two sources, we identified relevant excerpts related to mapping the risks of a system's use, minimizing its negative impacts while maximizing its benefits, and communicating these aspects to various stakeholders. Finally, from the ISO 42001, we identified excerpts on documenting resources for AI systems (data, tools, systems, computing, and human resources), assessing impacts on individuals, groups of individuals, and society, and reporting AI system details to stakeholders. 

Having these excerpts from previous literature at hand, we conducted 30-minute semi-structured interviews with 2 AI compliance experts to jointly annotate the collected excerpts (i.e., the EU AI Act, the NIST's AI RMF, and the ISO 42001) with high-level topics. We then jointly organized the topics in a thematic and hierarchical manner, resulting in the creation of the first version of the template (Figure~\ref{fig:methodology}, Step 1, V1). Through this joint exercise, we surfaced two main design requirements for an impact assessment report template: 
\begin{description}
    \item \emph{R1: Complete.} The report template should provide the necessary minimum information to complete an impact assessment in line with selected frameworks \cite{selbst2021algorithmicImpact, skoric2023assessmentCriteria}. The scope of this information must be carefully selected to enable consistent reporting, evaluation, and comparison of various AI systems.
    \item \emph{R2: Broad.} The report template should address all AI system components to identify, evaluate, and mitigate broad socio-technical impacts associated with this system's use. This approach allows for, firstly, identifying specific impacts related to individual system components, such as data sources or algorithms; and secondly, systematically broadening the scope to include other risks.
\end{description}

\subsection{Co-designing the Impact Assessment Report Template With AI Practitioners and AI Compliance Experts Through an Iterative Process}

After designing the first version of the template, we conducted a series of individual, 30-minute co-design sessions with 14 AI practitioners and 6 AI compliance experts (Figure \ref{fig:methodology}, Step 2). These sessions aimed to ensure the template provided sufficient information for impact assessment per the EU AI Act, NIST's AI RMF, and ISO 42001, and was usable by various roles in the assessment process.

\smallskip
\noindent\textbf{Participants.}
We aimed to achieve a diverse participant sample using snowball sampling, a method where existing study participants recruit future participants from among their acquaintances. We began by identifying 6 initial participants through an internal mailing list at a large tech company, seeking individuals familiar with the EU AI Act and actively involved in developing or evaluating at least one ongoing AI project during the time of the planned interviews. These participants were then asked to refer additional participants from their networks, thus expanding the sample size through successive referrals. We recruited a total of 20 participants (13 male, 7 female, with a median age of 34 years old) representing a variety of AI practitioners such as researchers (9), engineers (2), managers (3), designers (1), and AI compliance experts (6), and potential end users of the template. Their expertise span across various areas including generative AI, deep learning, AI standardization, and human rights. AI practitioners offered practical examples on navigating regulatory challenges in their projects  while maintaining innovation. Compliance experts contributed their knowledge of existing regulations, insights from ongoing regulatory works, and lessons learned from reviewing AI system uses. Table \ref{tab:codesign_demographics} provides an overview of the participants' demographics and relevant experience.

\smallskip
\noindent\textbf{Procedure.}
Before the session, we emailed participants with a demographic survey, a brief description of the session goals, the current iteration of the template, and summaries of the EU AI Act, NIST AI RMF, and ISO 42001 (Appendix A). We asked participants to familiarize themselves with these materials and optionally provide preliminary feedback by reviewing and annotating the template. This ensured participants understood the key regulatory frameworks, leading to more informed discussions and template revisions.

During the session, we first asked participants to evaluate the overall structure and design of the template. We then encouraged them to provide detailed comments on each section, focusing on any critical information they felt was missing or underrepresented. For compliance experts, we specifically sought to identify potential discrepancies in the template for missing information required for impact assessment in current AI regulations. We then asked all participants if they would use the template for their AI systems, and if so, how they would populate it. Finally, we discussed what features could be added to make the template more accessible to users with different roles.

We conducted individual sessions with 3 to 6 participants each, based on research suggesting that the first five participants can identify over 80\% of usability issues \cite{nielsen1993}. After each session, we summarized the prevalent issues and revised the template accordingly. The revised template was then tested in co-design sessions with the next batch of participants. Initially, we started the sessions exclusively with AI practitioners and included both AI practitioners and compliance experts after the first iteration.

By the time we produced the fifth version of the template (see iteration descriptions in Appendix B), we had gathered enough information from the co-design activities with AI practitioners and AI compliance experts that no new significant insights or usability issues were emerging. This indicates that the template had been refined to the point where additional testing was unlikely to yield further meaningful improvements. Following the approach of prior co-design studies \cite{Hupont2024} and cyclical action research \cite{eccola2021}, we concluded the template iterations at this stage and moved on to conducting a user study. 

\smallskip
\noindent\textbf{Analysis.}
After each co-design session, two authors conducted an inductive thematic analysis (bottom-up) of the session's transcripts, following established coding methodologies~\cite{saldana2015coding, miles1994qualitative, mcdonald2019reliability}. The authors used the Figma platform \cite{figma} to capture participants' feedback in sticky notes, and collaboratively created themes based on these notes. They discussed and resolved any disagreements that arose during the analysis to derive a list of usability issues to be addressed in the next iteration of the template.

These co-design sessions with AI practitioners and compliance experts surfaced two additional design requirements, which led to the design of a template in which no further refinements were deemed necessary (Figure~\ref{fig:methodology}, Step 2, V5):
\begin{description}
    \item \emph{R3: Adaptable to uses.} The report template should apply to a wide range of AI systems, their possible uses, and their application domains. It should also offer the flexibility to modify, or omit sections that may not apply to specific systems or their development stages, allowing for a more accurate and relevant evaluation.
    \item \emph{R4: Adaptable to different roles.} The report template should have a clear, self-contained structure with understandable sections, headings and subheadings. This allows users with different roles and levels of expertise with AI systems to effectively use it and participate in the impact assessment process.
\end{description}

\renewcommand{\arraystretch}{0.92}
\setlength{\tabcolsep}{2pt}
\begin{table*}[t!]
    \tiny
    \centering
    \caption{Demographics of AI compliance experts (E1-E2) who participated in semi-structured interviews and demographics of AI practitioners (P1-P14) and AI compliance experts (E3-E8) who participated in co-design sessions of the impact assessment report template. AI compliance experts highlighted that the current templates lacks scaffolding elements understandable to various roles and often miss the effects on individuals outside the direct users or subjects. AI practitioners recognized the need for these elements but struggle to define impacts on stakeholders beyond direct users, such as organizations, individuals, and societies, which are frequently referenced in regulations.}
    \label{tab:codesign_demographics}
    \begin{tabular}{p{2cm} p{0.7cm} p{1cm} p{1cm} p{1cm} p{3cm} p{1.7cm} p{2cm}}
    \toprule
    \textbf{Template version} & \textbf{ID} & \textbf{Gender} & \textbf{Age}  & \textbf{Education}  & \textbf{Expertise} & \textbf{Yrs of expr. in AI} & \textbf{Role}\\ \midrule
    
    \multirow{2}{*}{V1} & E1 & Female &	39 &	M.L. &	AI governance &	4 &	compliance expert \\
                        & E2 & Male &	43 &	MSc &		AI procurement &	5 &	compliance expert \\ 
    \midrule
    \multirow{3}{*}{V2} & P1 & Female & 33 & PhD &  deep learning & 5 & researcher \\
                        & P2 & Male &	30 &	PhD &		machine learning &	5 &	researcher \\
                        & P3 &	Male &	33 &	PhD &		machine learning &	5 &	researcher \\ 
                        & P4 &	Male &	32 &	PhD &		machine learning, NLP &	7 &	researcher \\ 
    \midrule
    \multirow{6}{*}{V3} & P5 &	Male &	35 &	PhD &	embedded machine learning &	5 &	manager \\
                        & P6 &	Male &	28 &	PhD &		mobile sensing &	6 &	researcher \\ 
                        & P7 &	Male &	37 &	PhD &		NLP &	10 &	researcher \\ 
                        & P8 &	Female &33 &	PhD &		machine learning &	2 &	engineer \\
                        & E3 & Female &	47 &	MA &		human rights impact assessment &	1 &	compliance expert \\
                        & E4 & Female &	33 &	MSc &		standardization &	7 &	compliance expert \\ 
    \midrule
    \multirow{5}{*}{V4} 
                        & P9 & Male &	59 &	BSc &		generative AI &	2 &	designer \\
                        & P10 &	Male &	32 &	MSc &		machine learning, NLP &	6 &	researcher \\
                        & P11 & Male &  27 &    MSc &       machine learning & 4 & researcher \\ 
                        & E5 & Male &	43 &	M.L. &		licensing &	1 &	compliance expert \\
                        & E6 & Female &	34 &	PhD &		standardization &	6 &	compliance expert \\ 
    \midrule
    \multirow{3}{*}{V5} & P12 &	Male &	43 &	PhD &		machine learning, computer vision &	5 &	researcher \\
                        & P13 &	Male &	40 &	PhD &		computer vision, bioinformatics &	5 &	researcher \\
                        & P14 & Female &26 &	MSc  & AI UX design  &	3 &	designer \\
                        & E7 & Female &	42 &	M.L &		AI case approval &	3 &	compliance expert \\
                        & E8 & Male &	50 &	MA &	AI procurement &	5 &	compliance expert \\ 
    \bottomrule
    \end{tabular}
\end{table*}

\subsection{Evaluating the Impact Assessment Report Template}
To then evaluate the final template, we first populated it with a real-world use of an AI system that was developed in the same large tech company and then conducted a user study to evaluate the populated report produced for this system's use with 8 AI practitioners and 5 AI compliance experts. 

\subsubsection{Populating the Impact Assessment Report Template.}
To populate the template, we employed a three-step semi-automatic method, which included soliciting AI practitioners responses and reviewing these responses with AI compliance experts (Figure \ref{fig:methodology}, Step 3). 

In the first step, we compiled a list of statements to gather responses from AI practitioners. To do so, we sourced statements from responsible AI guidelines \cite{ALTAI_2020,raiCrafting}, documentation standards \cite{selbst2021algorithmicImpact, gebru2021datasheets, holland2018dataset, bender2018data, mitchell2019model, sokol2020explanations, raji2020accountabilityGap}, checklists and impact assessment questionnaires \cite{madaio2020co, Golpayegani2023Risk, skoric2023assessmentCriteria, equalAI2023nist}. Next, we reviewed these statements, linking them to the relevant excerpts from the EU AI Act, the NIST AI RMF, and the ISO 42001, and grouping similar ones together. This process resulted in a list of 32 statements grounded in regulations and best responsible AI practices, designed to systematically gather information about the system's use, components, and data, team involvement, and the associated risks, mitigations, and benefits (Appendix C). They also directly map to the sections of our template.

In the second step, we reached out to 2 AI practitioners---a researcher and a designer---who had contributed to the development of an AI-based meeting companion app. We asked them to provide responses to these 32 statements.

In the third step, two authors manually parsed the practitioners' responses, placing them in the template to complete the report (Figure \ref{fig:final_template}). We then consulted 2 AI compliance experts, whom we had interviewed during the requirement elicitation phase, to review the report. They marked any legislative inconsistencies in the sections reporting risks and any factual inconsistencies in the sections reporting mitigation strategies and benefits. Upon inspecting the marked report, we found that the experts agreed on their assessments, and the report did not contain any inconsistencies. 

In addition to the report, we included a baseline condition to compare against our final template. For the baseline, we derived a generic impact assessment template for AI system use, based on the typical structure of existing impact assessment reports identified by \citeauthor{stahl2023systematicReview} (\citeyear{stahl2023systematicReview}). This generic template consisted of three sections: the first, ``Intended Use'', included a description of the use (covering the technology and application area) and stakeholders affected by the use; the second,``Risks'', covered human rights risks, ethical risks, data protection and privacy risks, safety risks, security risks, and environmental risks; and the third, ``Mitigation measures'', detailed technical and organizational measures for mitigating risks. 

The baseline content was the same as the final report except for describing intended use through full sentences, reorganizing risks and mitigations under different categories, and lacking the enumeration of benefits. Figures \ref{fig:meeting_companion} and Figure \ref{fig:meeting_companion_baseline} in Appendix D present the populated final impact assessment report and the baseline used in the user study.

\subsubsection{Conducting a User Study for Evaluating the Populated Impact Assessment Report.}
We conducted a user study in the form of semi-structured 30-minute interviews with 8 AI practitioners from the same large tech company and 5 AI compliance experts from both industry and academia (Figure \ref{fig:methodology}, Step 4). Specifically, our evaluation ought to answer four questions:

\begin{description}
    \item \emph{Q1:} To what extent does the report's content contain all necessary minimum information for conducting impact assessments in line with AI regulations (R1)?
    \item \emph{Q2:} To what extent does the report's content address all AI system components to identify, evaluate, and mitigate socio-technical risks of the system's use (R2)?
    \item \emph{Q3:} To what extent is the report's template adaptable to different AI system uses (R3)?
    \item \emph{Q4:} To what extent is the report's template adaptable to different roles (R4)?
\end{description}

\noindent\textbf{Participants.} We recruited 8 new AI practitioners working in the tech industry, with various roles and expertise, including computer vision engineers and researchers in deep learning, different from those who participated in the co-design sessions. Additionally, 5 new AI compliance experts took part in the study, who work in industry (3) and academia (2). Table \ref{tab:userstudy_demographics} summarizes participants' demographics. 
\smallskip

\noindent\textbf{Procedure.} Before conducting the interviews, we sent an email to all participants. This email included a brief description  of the study and a short survey focusing on demographics. The survey solicited details about the participants' age, area of expertise, professional role, and years of experience in developing AI systems (Appendix E). Furthermore, we attached brief overviews of the EU AI Act, the NIST AI RMF and the ISO 42001 (Appendix A). We requested the participants familiarize themselves with these documents to ensure a thorough evaluation of the template for alignment and more focused discussions during the interviews. Our organization gave its approval for the study. Again, we maintained data anonymity, removed personal identifiers, and restricted data access to the research team only.

During the interviews, we presented participants with both the populated final impact assessment report (Figure \ref{fig:meeting_companion}, Appendix C) and the baseline (Figure \ref{fig:meeting_companion_baseline}, Appendix C) for the AI-based meeting companion. Participants read each report for up to 15 minutes, alternating between them to avoid learning effects. After each reading, participants rated 12 statements (Figure \ref{fig:quantitative_results} S1-S12) related to the four design requirements identified from the co-design sessions, using a Likert scale from 1 (strong disagreement) to 5 (strong agreement). We then asked participants about their preferences, dislikes, and how they would adapt both report templates to their own work. Finally, two authors transcribed the interviews and conducted an inductive thematic analysis.
\smallskip

\noindent\textbf{Analysis.} First, we computed the average Likert scale score for each statement (Figure \ref{fig:quantitative_results}) for both the final report and the baseline. Second, two authors conducted an inductive thematic analysis (bottom-up) of the interview transcripts, following established coding methodologies~\cite{saldana2015coding, miles1994qualitative, mcdonald2019reliability}. The transcripts covered how the report's content supports impact assessment, how usable each template is, and any other preferences or dislikes.

The authors used the Figma platform \cite{figma} to capture participants' answers in sticky notes, and collaboratively created themes based on these notes. They discussed and resolved any disagreements that arose during the analysis process. Each theme included quotes from at least two participants, signifying that data saturation was reached~\cite{guest2006many}. 
\section{The Impact Assessment Report Template}

\begin{figure}[t!]
  \centering
  \includegraphics[width=\columnwidth]{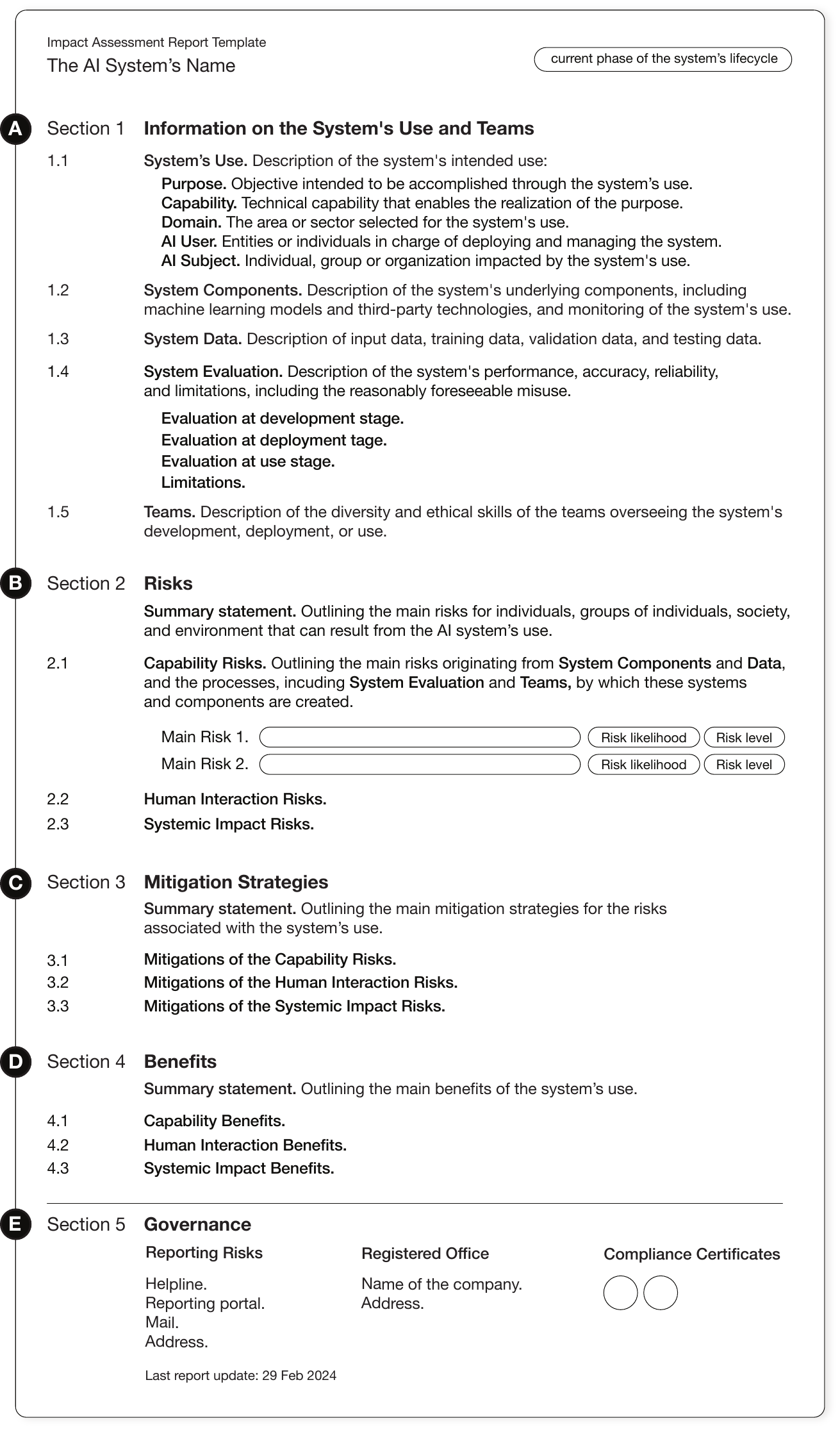}
  \caption{The final template for an Impact Assessment Report. Section 1 provides information on the system's use, components, data, evaluation, and teams; Section 2 lists potential risks; Section 3 lists mitigation strategies; Section 4 outlines the anticipated benefits from the system's use; and Section 5 outlines information about reporting mechanisms and who is responsible for the governance of the use.}
  \label{fig:final_template}
\end{figure}
\smallskip

The final template organizes information into five modular sections (Figure \ref{fig:final_template}A-E), allowing to add, remove or rearrange sections as needed (addressing R3):

\smallskip
\noindent\textbf{Section 1: Information on the System's Use and Teams.} 
The section begins with a clear description of the system's intended use (Figure \ref{fig:final_template}A), including its intended purpose, the technical capabilities that enable this use, the area or sector chosen for its application (including its geographic and temporal extent), any natural persons and groups likely to be affected by this use, and any natural or legal persons who have authority over it. This description mirrors the five risk assessment components identified by~\citet{Golpayegani2023Risk}: \emph{purpose}, \emph{capability}, \emph{domain}, \emph{user}, and \emph{subject} that provide information necessary for conducting the risk assessment based on the EU AI Act (addressing R1). Users could specify these components using external dictionaries, such as the Vocabulary of AI Risks \cite{VAIR2023Framework}, which contains pre-defined descriptions of these concepts directly sourced from the Act.

Following that, the section contains two subsections related to system's underlying components (including its machine learning models, third-party technologies, and mechanisms for system's monitoring) and data (addressing R2). Afterwards, it contains a subsection related to system's evaluation at different stages of the system's lifecycle (e.g., development, deployment, and use as per AI RMF) including description of the system's performance, accuracy, reliability, and limitations (e.g., the reasonably foreseeable misuse). The section concludes by disclosing the diversity of the teams overseeing the system's use at different stages. 

Additional subsections could be included to address specific regulatory requirements of different regions or to provide more contextual information about the system's use, such as previous experiences with the deployment of similar systems (addressing R3 and R4).

\smallskip
\noindent\textbf{Section 2: Risks.} 
This section lists potential risks (Figure \ref{fig:final_template}B) associated with putting the system into use (addressing R2). To begin with, it presents a brief summary statement outlining the main risks that may undermine safety, rule of law, fundamental rights, health, environment, and democracy. It then lists all the specific risks of system's use grouped into risks stemming from the \emph{system's capability} (i.e., the technical components of the AI system), \emph{human interaction} (i.e., experiences of people interacting with the AI system), and \emph{systemic impact} (i.e., societal, economic, and environmental impacts of the AI system's use). This grouping aligns with a three-layered framework for evaluating sociotechnical harms by Google DeepMind \cite{weidinger2023sociotechnical}. It was selected because AI practitioners found it more approachable (addressing R4), based on feedback indicating difficulties in envisioning realistic risks across stakeholders (i.e., organizations, individuals, groups of individuals, and societies) listed in the EU AI Act, AI RMF, and ISO 42001. For each risk listed in the section, we have added to the template two placeholders specifically for `Risk likelihood' and `Risk level'. This encourages users of the template to categorize each risk's impact by estimating its probability and potential impact as high, medium, low, respectively, and record these assessments in the provided placeholders. This facilitates impact assessment per ISO 42001 (addressing R1) and allows estimation of the overall risk level score of the use as a combination of the risk likelihoods and risk levels.

\smallskip
\noindent\textbf{Section 3: Mitigation Strategies.} 
This section lists mitigation strategies (Figure \ref{fig:final_template}C) designed to address and minimize the previously identified risks of system's use (addressing R2). The section provides a brief summary statement highlighting the main mitigation strategies for the system's use, and then a list of mitigation strategies for each risk group identified in Section 2 (Figure \ref{fig:final_template}B).

\smallskip
\noindent\textbf{Section 4: Benefits.} 
This section outlines the anticipated benefits and positive impacts resulting from the system's use (Figure \ref{fig:final_template}D). First, the section provides a brief summary statement covering direct and indirect advantages for users, organizations, and society, emphasizing potential long-term positive effects. It then lists all the benefits of the system's use grouped into themes, such as improving access to quality education. By doing so, the section balances the discussion of risks and fosters a comprehensive understanding of the overall impact of the system's use. 

\smallskip
\noindent\textbf{Section 5: Governance.} 
This final section (Figure \ref{fig:final_template}E) outlines information about reporting mechanisms (e.g., dedicated email, phone number, the registered office) and who is responsible for the governance of the use (e.g., European conformity marking and other compliance certifications issued so far for the use; addressing R2).

The template provides clear section and subsection titles to help users quickly identify the content they are looking for (addressing R4). Having public interpretability in view, we also applied best practices from Information Visualization such as consistent formatting style throughout the template, including headings, sans-serif fonts, and ample white spacing \cite{visWhatWorks2021}.

The resulting template aligns with regulatory guidelines \cite{EUACT2021, nist2023aiRisk} while offering an alternative approach to current reporting practices \cite{stahl2023systematicReview}. Instead of broadly considering the AI models or systems, we have focused our template on the system's intended use for which prevalent documentation is typically limited to a brief textual description without a standardized format \cite{intendedUse2021}. 

\section{User Study of the Impact Assessment Report Template}

\renewcommand{\arraystretch}{1.1}
\setlength{\tabcolsep}{2pt}
\begin{table*}[t!]
    \tiny
    \centering
    \caption{Demographics of AI practitioners (P15-P22) and AI compliance experts (E9-E13) from industry and academia (marked with *) who participated in the template evaluation user study.}
    \label{tab:userstudy_demographics}
    \begin{tabular}{p{2.2cm} p{0.7cm} p{1cm} p{1cm} p{1cm} p{3cm} p{1.7cm} p{2cm}}
    \toprule
    \textbf{Group} & \textbf{ID} & \textbf{Gender} & \textbf{Age}  & \textbf{Education} & \textbf{Expertise} & \textbf{Yrs of expr. in AI} & \textbf{Role}\\ \midrule
    
    \multirow{8}{*}{AI Practitioners} & P1 & Female & 33 & PhD & deep learning & 5 & researcher \\
                        & P2 & Male &	31 &	PhD  &	machine learning &	5 &	researcher \\
                        & P3 &	Male &	33 &	PhD &	machine learning &	5 &	researcher \\ 
                        & P4 &	Male &	32 &	PhD &	machine learning &	7 &	researcher \\ 
                        & P5 &	Male &	36 &	PhD &	embedded machine learning &	5 &	manager \\
                        & P6 &	Male &	27 &	PhD &	mobile sensing &	6 &	researcher \\ 
                        & P7 &	Male &	37 &	PhD &	NLP &	10 &	researcher \\ 
                        & P8 &	Female & 33 &	PhD &	machine learning &	2 &	engineer \\ \midrule            
    \multirow{5}{*}{AI Compliance Experts} 
                        & E9 & Female &	47 &	MA &	human rights law &	1 &	manager \\
                        & E10 & Female &	33 &	MSc &	standardization &	7 &	executive advisor \\ 
                        & E11* & Female &	25 &	PhD &	impact assessment &	5 & researcher\\
                        & E12* & Female &	32 &	MSc &	policy making, content moderation &	2 &	researcher \\ 
                        & E13 & Male &	44 &	MSc &	audit, risk assessment & 2 &	manager \\     
    \bottomrule
    \end{tabular}
\end{table*}

\begin{figure}[t!]
  \centering
\includegraphics[width=\columnwidth]{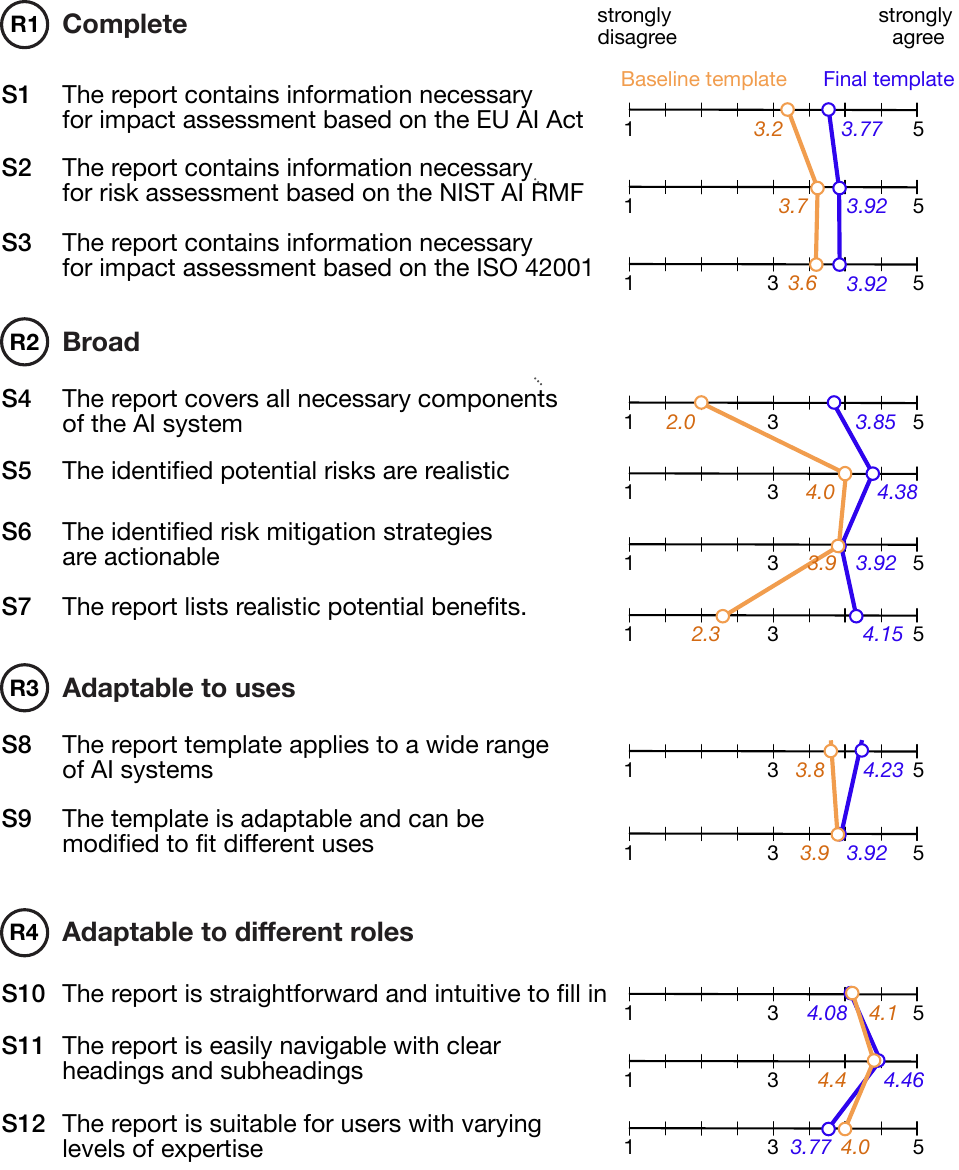}
  \caption{Ratings on twelve statements regarding the four requirements -- R1: Complete, R2: Broad, R3: Adaptable to uses, and R4: Adaptable to different roles -- for both the baseline template and the final template. Participants found that the final report provided more complete information for impact assessments (R1) and addressed all AI system components and impacts more broadly than the baseline (R2).}
  \label{fig:quantitative_results}
\end{figure}

We report the quantitative results with the qualitative results, which add the participants' perspectives on what could be improved in the next co-design iterations.

\subsubsection{Quantitative results.}
Participants found that the final report provided more complete information for conducting impact assessments than the baseline across all three frameworks (Figure \ref{fig:quantitative_results}, R1). The report received the highest ratings for alignment with the NIST AI RMF and ISO 42001, and the lowest for the EU AI Act. Participants also found that the final report provided a broader scope in addressing all AI system components and the impacts of their use compared to the baseline (Figure \ref{fig:quantitative_results}, R2). In terms of adaptability to different uses (Figure \ref{fig:quantitative_results}, R3), participants found the final report slightly more applicable to a wide range of AI systems than the baseline. Regarding adaptability to different roles (Figure \ref{fig:quantitative_results}, R4), all participants rated both the final template and the baseline as very straightforward and intuitive to complete, easy to navigate with clear headings and subheadings, and suitable for users with varying roles.

\subsubsection{Qualitative results.}
AI practitioners found the final template and its sections helpful. For example, P6 stated that \emph{``I really like it [the template] covers various aspects of the system like its components and data''}. Practitioners also praised the template's simplicity. P3 stressed they \emph{``like its simplicity as it helps me deal with the complexity of the AI system''}, for example by scoping well the intended use of the system, which P7 summarized as \emph{``you see it [five-format component], and it sticks with you''}. All 8 AI practitioners praised the idea of grounding the report's content in the EU AI Act articles. For example, P8 stated that \emph{``these risks are citing the specific sections of the Act [...] It's definitely enough information to convince me that it [the meeting companion] is high risk''}. This was more evident when our participants commented about the risks section of the report. Specifically, P1 mentioned that \emph{``the risk section is a good start for understanding how it [the meeting companion] can be misused''}. However, AI practitioners also saw a number of improvements. Six respondents highlighted the need for automated tools to populate the report. E10 stated that \emph{``it is s very much about discovering risks and I strongly believe that we should use automated tools and GenAI to aid in that process to prompt with the right kind of responses''}.

AI compliance experts have confirmed that the final template effectively addresses two aspects of integrating and managing AI systems within organizational processes. First, the final template promotes the integration of AI-driven processes with existing, well-established organizational procedures. As stated by E13, \emph{``the approach with the evaluation during development, deployment and use is quite familiar to anyone really involved in product development. This feedback must have been given by developers and I agree - completing the template should not be an additional burdensome process, but something that is integrated into what we are already used to doing''}. Second, by advocating for the blending of multiple taxonomies and frameworks, the template ensures a thorough and comprehensive approach to risk assessment. This method helps prevent the oversight of potential risks by discouraging a narrow focus on single frameworks. As summarized by E10, \emph{``I see your intention not to focus on only one risk taxonomy. If you want to make sure that you are covering everything, it is very good to start with one taxonomy and blend in taxonomies from other frameworks. There are no clear boundaries between risks; they overlap and create different categories across frameworks. In reality, these frameworks interact and mix to some extent.''} To see how, E12 explained how the same term ``systemic risk'' has multiple meaning based on the regulation: \emph{``very large online platforms and search engines, e.g., providing AI-based recommender systems, need to perform a systemic risk assessment under the Digital Services Act, which specifies four categories of such risks [the dissemination of illegal content, negative effects for the fundamental rights; negative effects for the civic discourse, electoral processes, and public security; and negative effects in relation to gender-based violence, the protection of public health and minors and serious negative consequences to the person’s physical and mental well-being]. However, systemic risk in the EU AI Act relates also to the general purpose AI models and refers to negative effects on public health, safety, public security, fundamental rights, or the society''}.

Both AI practitioners and AI compliance experts stressed the importance of the inclusion of expert oversight to elevate the quality and reliability of the assessments carried out jointly by these stakeholders. E9, AI compliance expert, echoed AI practitioners' perspectives: \emph{``the biggest challenge is ensuring teams have the necessary skills and knowledge to complete the assessment. Good written guidelines and preferably some expert guidance and oversight would lead to better results''}. Participants also envisioned using the template not only at the pre-deployment stage right before the compliance but also as a tool to guide the design stage of AI uses. P7, researcher, noted that the template \emph{``suits all roles and summarizes the design decisions made so far. It has a different structure than typical compliance reports, where each section has a topic-specific focus and is relevant only for certain experts''}. 
E12, a compliance expert, suggested that \emph{``the template can be completed not only at the development stage but also every time the socio-technical context changes''}. To illustrate this point, they gave the example of the national system that evaluates the chances of specific groups in the labor market, which was not usable during the COVID-19 pandemic, and for which similar risks had not been foreseen in the limitations of the system.
\section{Discussion}
\label{sec:discussion}

We contextualize our template within prior literature, then discuss its implications, limitations, and future research.

\subsection{Differences from Existing Templates}
\label{subsec:contextualize_template}
We compared our template with the Algorithmic Impact Assessment template from NIST~\cite{equalAI2023nist}, Microsoft's Responsible AI Impact Assessment Template~\cite{microsoft2022raiImpactAssessmentTemplate}, Credo AI's Standardized Risk Profile~\cite{sherman2023riskProfiles}, and the Algorithmic Impact Assessment from the Ada Lovelace Institute~\cite{ada_lovelace}, and identified four key differences. First, our template enhances cross-company comparisons by aligning closely with legislation rather than relying on framework-specific biases or company-specific risks. Second, unlike existing templates which focus heavily on risks, ours equally emphasizes both risks and benefits. Third, there is variation in the extent of guidance provided for completing the templates. NIST's~\cite{equalAI2023nist}, Microsoft's \cite{microsoft2022raiImpactAssessmentGuide}, and the Ada Lovelace Institute's~\cite{ada_lovelace} offer guidebooks on how to produce a report. Rather than leaving stakeholders with a blank report and a guide, we offer a guidebook through our 32 simple statements that alleviate potential anxiety associated with an empty page and facilitate the identification of risks and benefits~\cite{reportProjectPage}. Finally, whereas other templates have separate sections for legal and compliance issues, ours integrates these aspects throughout.

\subsection{Theoretical and Practical Implications}
\label{subsec:theoretical}

\smallskip
\noindent\textbf{Embedding AI governance in impact assessment reports.} 
We introduced a standardized template that not only facilitates risk assessment in accordance with the EU AI Act, the NIST's AI RMF, the ISO 42001 but also serves as a model for integrating regulatory considerations into AIIAs. Our work contributes to co-designing and validating improved AIIAs and the processes and workflows surrounding them \cite{skoric2023assessmentCriteria}. This includes developing new tools and methodologies that are practical and adaptable to enhance the quality and effectiveness of impact assessments in the AI context. They provide a comprehensive and practical framework for companies to navigate the complexities of AI regulation, ensuring both compliance and ethical responsibility in AI development and implementation.

\smallskip
\noindent\textbf{Facilitating contextual evaluation.} Finding the right balance between making impact assessment template general enough to apply to various AI systems and specific enough to provide meaningful assessments for each unique system's use is a significant challenge \cite{stahl2023systematicReview}. We partly addressed this challenge by adding specific subsections into the template. These subsections cover the five components comprehensively describing the system's use, the three key stakeholders, and the three stages of the system's lifecycle. By doing so, we assist stakeholders in systematically documenting the intended use for which the system was built, ensuring that it meets performance and safety criteria across different situations \cite{johnson2023classroomStudy}.

\smallskip
\noindent\textbf{Improving stakeholder engagement.} Our report template can be utilized by individual team members or as a group. For teams that are new to impact assessments, working through this template can be an educational experience that helps in building understanding and skills related to responsible AI practices. Future work on improving stakeholder engagement should explore alternative interactive tools that effectively balance various methods of eliciting system information and envisioning system impacts, such as divergent and convergent thinking styles \cite{selbst2021algorithmicImpact}.

\subsection{Limitations and Future Work}
\label{subsec:limitations}

\noindent\textbf{Generalizability.}
Our results are based on a pool of study participants who were familiar with the EU AI Act, the NIST framework and ISO standards. Including a broader range of roles beyond managers, designers, and researchers may yield different results. Future studies should include participants with different levels of knowledge about AI regulations. They should also evaluate how well impact assessment reports explain the risks and benefits of AI to everyone and find ways to help people understand important AI regulations and laws. 

\smallskip
\noindent\textbf{Propagating biases in impact assessment.} AIIAs, like other responsible AI tools, have inherent biases from their design choices, such as excluding potential users \cite{Moss2021assemblingAccountability}.
The quality of our report template depends on the accuracy and completeness of user-provided information. Using biased or incomplete data can lead to incorrect assessments, making problems seem smaller than they really are because people might be afraid to report negative impacts. To fix this, future research should include external perspectives by holding workshops, involving independent experts and marginalized groups, and regularly checking the results with new team members \cite{raji2020accountabilityGap, ada_lovelace}.

\smallskip
\noindent\textbf{Automated tools pre-populating the template.} 
Automated tools can help make gathering data for templates easier and less prone to mistakes. Large Language Models (LLMs) can help fill out impact assessment reports by generating lists of AI users and subjects \cite{bucinca2023harmGeneration}, identifying intended and unintended uses \cite{farsight2024, herdel2024exploregenlargelanguagemodels}, and listing potential risks and benefits \cite{constantinides2024_risks_benefits, velazquez2024}. \citet{AIDesign2024} have recently proposed  a semi-automatic system that collects input from stakeholders about an AI system's use , uses LLMs to find additional risks, mitigation strategies, and benefits, and pre-fills reports for experts to review.

\smallskip
\noindent\textbf{Responsible by Design.} AIIAs are usually conducted at the pre-deployment stage for legal compliance. However, after our co-design process and user studies which began with compliance as a focal point, we learned that reports should be updated not only when lifecycle stages change (e.g., from design to development), but also when the socio-technical context shifts. This underscores the challenge of balancing a fixed impact assessment template with the significant learning benefits users gain from engaging with it to ``think differently'' about AI. Without clear updates, users may lack the critical thinking and ethical deliberation needed, ultimately reducing empathy for those impacted by AI uses.
\section{Conclusion}
We developed an impact assessment template with input from 16 AI practitioners and 6 compliance experts, designed to align with standards such as the EU AI Act, NIST AI RMF, and ISO 42001. A user study involving other 8 company AI practitioners and 5 compliance experts confirmed that our template effectively captures AI system impacts, serving as a starting point for navigating regulatory compliance and fostering responsible design.

\newpage
\section{Researcher Positionality Statement}
In this study, we, the authors, are based in United Kingdom and engage predominantly in industry and academic research during the 21\textsuperscript{st} century. Our team includes two women and two men from Southern, Eastern, and Central Europe, representing a variety of ethnic backgrounds. Our collective expertise spans several fields such as human-computer interaction, ubiquitous computing, software engineering, artificial intelligence, natural language processing, data visualization, and digital humanities. Our positionality may influence the inherent subjectivity in formulating our research, choosing our methods, structuring our co-design and evaluation sessions, interpreting and analyzing data, and addressing the needs of study participants in future iterations of the template \cite{frluckaj2022gender}.

As researchers in a predominantly Western setting, we understand the critical importance of broadening the scope of perspectives in our research, particularly to include voices from outside academic and industry spheres. We are committed to promoting future research that is conducted by and with individuals from a wide range of backgrounds, especially those with personal experiences of the impacts of AI systems. This inclusive approach will deepen our understanding and help us develop research methods that are truly responsive to the needs of diverse roles and often underrepresented communities.

%%% Bibliography
\bibliography{aaai24}

\begin{thebibliography}{66}
\providecommand{\natexlab}[1]{#1}

\bibitem[{{Ada Lovelace Institute}(2022)}]{ada_lovelace}
{Ada Lovelace Institute}. 2022.
\newblock Algorithmic Impact Assessment: AIA Template.
\newblock Available at: www.adalovelaceinstitute.org/resource/aia-template.

\bibitem[{Ballard, Chappell, and Kennedy(2019)}]{judgmentCall_2019}
Ballard, S.; Chappell, K.~M.; and Kennedy, K. 2019.
\newblock {Judgment Call the Game: Using Value Sensitive Design and Design Fiction to Surface Ethical Concerns Related to Technology}.
\newblock In \emph{Proceedings of the ACM on Designing Interactive Systems Conference}, 421–433.

\bibitem[{Barnett and Diakopoulos(2022)}]{Barnett2022}
Barnett, J.; and Diakopoulos, N. 2022.
\newblock Crowdsourcing Impacts: Exploring the Utility of Crowds for Anticipating Societal Impacts of Algorithmic Decision Making.
\newblock In \emph{Proceedings of the AAAI/ACM Conference on AI, Ethics, and Society}, 56–67.

\bibitem[{Bender and Friedman(2018)}]{bender2018data}
Bender, E.~M.; and Friedman, B. 2018.
\newblock {Data Statements for Natural Language Processing: Toward Mitigating System Bias and Enabling Better Science}.
\newblock \emph{Transactions of the Association for Computational Linguistics}, 6: 587--604.

\bibitem[{Bogucka et~al.(2024{\natexlab{a}})Bogucka, Constantinides, Šćepanović, and Quercia}]{AIDesign2024}
Bogucka, E.; Constantinides, M.; Šćepanović, S.; and Quercia, D. 2024{\natexlab{a}}.
\newblock AI Design: A Responsible AI Framework for Impact Assessment Reports.
\newblock \emph{Internet Computing}.

\bibitem[{Bogucka et~al.(2024{\natexlab{b}})Bogucka, Constantinides, Šćepanović, and Quercia}]{reportProjectPage}
Bogucka, E.; Constantinides, M.; Šćepanović, S.; and Quercia, D. 2024{\natexlab{b}}.
\newblock Co-designing an AI Impact Assessment Report Template with AI Practitioners and AI Compliance Experts - Supplementary Materials.
\newblock Available at: https://social-dynamics.net/impact-assessment.

\bibitem[{Buçinca et~al.(2023)Buçinca, Pham, Jakesch, Ribeiro, Olteanu, and Amershi}]{bucinca2023harmGeneration}
Buçinca, Z.; Pham, C.~M.; Jakesch, M.; Ribeiro, M.~T.; Olteanu, A.; and Amershi, S. 2023.
\newblock AHA!: Facilitating AI Impact Assessment by Generating Examples of Harms.
\newblock arXiv:2306.03280.

\bibitem[{Constantinides et~al.(2024{\natexlab{a}})Constantinides, Bogucka, Quercia, Kallio, and Tahaei}]{raiCrafting}
Constantinides, M.; Bogucka, E.; Quercia, D.; Kallio, S.; and Tahaei, M. 2024{\natexlab{a}}.
\newblock RAI Guidelines: Method for Generating Responsible AI Guidelines Grounded in Regulations and Usable by (Non-)Technical Roles.
\newblock In \emph{Proceedings of the ACM on Human-Computer Interaction}, CSCW, 1--28.

\bibitem[{Constantinides et~al.(2024{\natexlab{b}})Constantinides, Bogucka, Scepanovic, and Quercia}]{constantinides2024_risks_benefits}
Constantinides, M.; Bogucka, E.; Scepanovic, S.; and Quercia, D. 2024{\natexlab{b}}.
\newblock Good Intentions, Risky Inventions: A Method for Assessing the Risks and Benefits of AI in Mobile and Wearable Uses.
\newblock In \emph{Proceedings of the ACM on Human-Computer Interaction}, volume~8, 1--30.

\bibitem[{De~Miguel~Velazquez et~al.(2024)De~Miguel~Velazquez, Šćepanović, Gvirtz, and Quercia}]{velazquez2024}
De~Miguel~Velazquez, J.; Šćepanović, S.; Gvirtz, A.; and Quercia, D. 2024.
\newblock Decoding Real-World AI Incidents.
\newblock \emph{IEEE Computer}.

\bibitem[{Diakopoulos et~al.(2016)Diakopoulos, Friedler, Arenas, Barocas, Hay, Howe, Jagadish, Unsworth, Sahuguet, Venkatasubramanian, Wilson, Yu, and Zevenbergen}]{diakopoulos2016socialImpactStatement}
Diakopoulos, N.; Friedler, S.; Arenas, M.; Barocas, S.; Hay, M.; Howe, B.; Jagadish, H.~V.; Unsworth, K.; Sahuguet, A.; Venkatasubramanian, S.; Wilson, C.; Yu, C.; and Zevenbergen, B. 2016.
\newblock Principles for Accountable Algorithms and a Social Impact Statement for Algorithms.
\newblock Available at: www.fatml.org/resources/principles-for-accountable-algorithms.

\bibitem[{Ehsan et~al.(2022)Ehsan, Singh, Metcalf, and Riedl}]{ehsan2022algorithmicImprint}
Ehsan, U.; Singh, R.; Metcalf, J.; and Riedl, M. 2022.
\newblock The Algorithmic Imprint.
\newblock In \emph{Proceedings of the ACM Conference on Fairness, Accountability, and Transparency}, 1305–1317.

\bibitem[{{European Comission}(2024)}]{EUACT2021}
{European Comission}. 2024.
\newblock Regulation of the European Parliament and of the Council on Laying Down Harmonised Rules on Artificial Intelligence and Amending Regulations (EC) No 300/2008, (EU) No 167/2013, (EU) No 168/2013, (EU) 2018/858, (EU) 2018/1139 and (EU) 2019/2144 and Directives 2014/90/EU, (EU) 2016/797 and (EU) 2020/1828 (Artificial Intelligence Act).
\newblock Available at: \url{www.europarl.europa.eu/doceo/document/TA-9-2024-0138-FNL-COR01_EN.pdf}.

\bibitem[{Figma(2024)}]{figma}
Figma. 2024.
\newblock {Figma: The Collaborative Interface Design Tool}.
\newblock Available at: \url{www.figma.com}.

\bibitem[{Franconeri et~al.(2021)Franconeri, Padilla, Shah, Zacks, and Hullman}]{visWhatWorks2021}
Franconeri, S.~L.; Padilla, L.~M.; Shah, P.; Zacks, J.~M.; and Hullman, J. 2021.
\newblock The Science of Visual Data Communication: What Works.
\newblock \emph{Psychological Science in the Public Interest}, 22(3): 110--161.

\bibitem[{Frluckaj et~al.(2022)Frluckaj, Dabbish, Widder, Qiu, and Herbsleb}]{frluckaj2022gender}
Frluckaj, H.; Dabbish, L.; Widder, D.~G.; Qiu, H.~S.; and Herbsleb, J. 2022.
\newblock Gender and Participation in Open Source Software Development.
\newblock In \emph{Proceedings of the ACM on Human-Computer Interaction}, volume~6, 1--31. ACM.

\bibitem[{{Fujitsu Research Center for AI Ethics}(2022)}]{fujitsu2022ethicalAssessment}
{Fujitsu Research Center for AI Ethics}. 2022.
\newblock AI Ethics Impact Assessment.
\newblock Technical report, Fujitsu.
\newblock Available at: www.fujitsu.com/global/about/research/technology/aiethics.

\bibitem[{Gaba et~al.(2024)Gaba, Kaufman, Chueng, Shvakel, Hall, Brun, and Bearfield}]{gaba2023visualDesign}
Gaba, A.; Kaufman, Z.; Chueng, J.; Shvakel, M.; Hall, K.~W.; Brun, Y.; and Bearfield, C.~X. 2024.
\newblock My Model is Unfair, Do People Even Care? Visual Design Affects Trust and Perceived Bias in Machine Learning.
\newblock \emph{IEEE Transactions on Visualization and Computer Graphics}, 30(1): 327--337.

\bibitem[{Gaumond and Régis(2023)}]{hria_2023}
Gaumond, E.; and Régis, C. 2023.
\newblock Assessing Impacts of AI on Human Rights: It’s Not Solely About Privacy and Nondiscrimination.
\newblock Available at: www.lawfaremedia.org/article/assessing-impacts-of-ai-on-human-rights-it-s-not-solely-about-privacy-and-nondiscrimination.

\bibitem[{Gebru et~al.(2021)Gebru, Morgenstern, Vecchione, Vaughan, Wallach, Daum\'{e}~III, and Crawford}]{gebru2021datasheets}
Gebru, T.; Morgenstern, J.; Vecchione, B.; Vaughan, J.~W.; Wallach, H.; Daum\'{e}~III, H.; and Crawford, K. 2021.
\newblock {Datasheets for Datasets}.
\newblock \emph{Communications of the ACM}, 64(12): 86–92.

\bibitem[{Golbin(2021)}]{pwc2021impactAssessment}
Golbin, I. 2021.
\newblock Algorithmic Impact Assessments: What Are They and Why Do You Need Them?
\newblock Available at: www.pwc.com/us/en/tech-effect/ai-analytics/algorithmic-impact-assessments.html.

\bibitem[{Golpayegani, Pandit, and Lewis(2023)}]{Golpayegani2023Risk}
Golpayegani, D.; Pandit, H.~J.; and Lewis, D. 2023.
\newblock To Be High-Risk, or Not To Be--Semantic Specifications and Implications of the AI Act's High-Risk AI Applications and Harmonised Standards.
\newblock In \emph{Proceedings of the ACM Conference on Fairness, Accountability, and Transparency}, 905--915.

\bibitem[{Golpaygani, Pandit, and Lewis(2023)}]{VAIR2023Framework}
Golpaygani, D.; Pandit, H.~J.; and Lewis, D. 2023.
\newblock VAIR: Vocabulary of AI Risks.
\newblock Available at: delaramglp.github.io/vair.

\bibitem[{Gomes~de Andrade and Kontschieder(2021)}]{Andrade2021}
Gomes~de Andrade, N.~N.; and Kontschieder, V. 2021.
\newblock AI Impact Assessment: A Policy Prototyping Experiment.
\newblock \emph{SSRN Electronic Journal}.

\bibitem[{Guest, Bunce, and Johnson(2006)}]{guest2006many}
Guest, G.; Bunce, A.; and Johnson, L. 2006.
\newblock How many interviews are enough? An experiment with data saturation and variability.
\newblock \emph{Field Methods}, 18(1): 59--82.

\bibitem[{Herdel et~al.(2024)Herdel, Šćepanović, Bogucka, and Quercia}]{herdel2024exploregenlargelanguagemodels}
Herdel, V.; Šćepanović, S.; Bogucka, E.; and Quercia, D. 2024.
\newblock ExploreGen: Large Language Models for Envisioning the Uses and Risks of AI Technologies.
\newblock arXiv:2407.12454.

\bibitem[{Holland et~al.(2020)Holland, Hosny, Newman, Joseph, and Chmielinski}]{holland2018dataset}
Holland, S.; Hosny, A.; Newman, S.; Joseph, J.; and Chmielinski, K. 2020.
\newblock The Dataset Nutrition Label: A Framework to Drive Higher Data Quality Standards.
\newblock In Hallinan, D.; Leenes, R.; Gutwirth, S.; and De~Hert, P., eds., \emph{Data Protection and Privacy}, chapter~1, 1--26. Hart Publishing.

\bibitem[{Hupont et~al.(2024)Hupont, Fernández-Llorca, Baldassarri, and Gómez}]{Hupont2024}
Hupont, I.; Fernández-Llorca, D.; Baldassarri, S.; and Gómez, E. 2024.
\newblock Use Case Cards: A Use Case Reporting Framework Inspired by the European AI Act.
\newblock \emph{Ethics and Information Technology}, 26(2).

\bibitem[{{ISO/IEC}(2023)}]{iso2023ManagementSystem}
{ISO/IEC}. 2023.
\newblock Information Technology -- {Artificial Intelligence} -- {Management System}.
\newblock Standard ISO/IEC 42001:2023, International Organization for Standardization.

\bibitem[{{ISO/IEC}(2025)}]{iso2025aiSystemImpactAssessment}
{ISO/IEC}. 2025.
\newblock Information Technology -- {Artificial Intelligence} -- {AI System Impact Assessment}.
\newblock Standard ISO/IEC DIS 42005, International Organization for Standardization.
\newblock Status: Under development.

\bibitem[{Janssen(2020)}]{fundamentalRightsList_2020}
Janssen, H.~L. 2020.
\newblock {An Approach for a Fundamental Rights Impact Assessment to Automated Decision-Making}.
\newblock \emph{International Data Privacy Law}, 10(1): 76--106.

\bibitem[{Johnson and Heidari(2023)}]{johnson2023classroomStudy}
Johnson, N.; and Heidari, H. 2023.
\newblock Assessing {AI} Impact Assessments: A Classroom Study.
\newblock In \emph{NeurIPS 2023 Workshop on Regulatable Machine Learning}.

\bibitem[{Madaio et~al.(2020)Madaio, Stark, Wortman~Vaughan, and Wallach}]{madaio2020co}
Madaio, M.~A.; Stark, L.; Wortman~Vaughan, J.; and Wallach, H. 2020.
\newblock {Co-designing Checklists to Understand Organizational Challenges and Opportunities Around Fairness in AI}.
\newblock In \emph{Proceedings of the ACM Conference on Human Factors in Computing Systems}, 1--14.

\bibitem[{Mantelero(2022)}]{mantelero2022}
Mantelero, A. 2022.
\newblock \emph{Beyond Data: Human Rights, Ethical and Social Impact Assessment in AI}.
\newblock T.M.C. Asser Press.

\bibitem[{McDonald, Schoenebeck, and Forte(2019)}]{mcdonald2019reliability}
McDonald, N.; Schoenebeck, S.; and Forte, A. 2019.
\newblock {Reliability and Inter-Rater Reliability in Qualitative Research: Norms and Guidelines for CSCW and HCI Practice}.
\newblock In \emph{Proceedings of the ACM on Human-Computer Interaction}, volume~3. ACM.

\bibitem[{Metcalf et~al.(2021)Metcalf, Moss, Watkins, Singh, and Elish}]{metcalf2021constructingImpacts}
Metcalf, J.; Moss, E.; Watkins, E.~A.; Singh, R.; and Elish, M.~C. 2021.
\newblock Algorithmic Impact Assessments and Accountability: The Co-Construction of Impacts.
\newblock In \emph{Proceedings of the ACM Conference on Fairness, Accountability, and Transparency}, 735–746.

\bibitem[{Microsoft(2022{\natexlab{a}})}]{microsoft2022raiImpactAssessmentGuide}
Microsoft. 2022{\natexlab{a}}.
\newblock Responsible AI Impact Assessment Guide.
\newblock Available at: blogs.microsoft.com/wp-content/uploads/prod/sites/5/2022/06/Microsoft-RAI-Impact-Assessment-Guide.pdf.

\bibitem[{Microsoft(2022{\natexlab{b}})}]{microsoft2022raiImpactAssessmentTemplate}
Microsoft. 2022{\natexlab{b}}.
\newblock Responsible AI Impact Assessment Template.
\newblock Available at: blogs.microsoft.com/wp-content/uploads/prod/sites/5/2022/06/Microsoft-RAI-Impact-Assessment-Template.pdf.

\bibitem[{Miles and Huberman(1994)}]{miles1994qualitative}
Miles, M.; and Huberman, M. 1994.
\newblock \emph{{Qualitative Data Analysis: A Methods Sourcebook}}.
\newblock Sage.

\bibitem[{Mitchell et~al.(2019)Mitchell, Wu, Zaldivar, Barnes, Vasserman, Hutchinson, Spitzer, Raji, and Gebru}]{mitchell2019model}
Mitchell, M.; Wu, S.; Zaldivar, A.; Barnes, P.; Vasserman, L.; Hutchinson, B.; Spitzer, E.; Raji, I.~D.; and Gebru, T. 2019.
\newblock {Model Cards for Model Reporting}.
\newblock In \emph{Proceedings of the ACM Conference on Fairness, Accountability, and Transparency}, 220–229.

\bibitem[{Moss et~al.(2021)Moss, Watkins, Singh, Elish, and Metcalf}]{Moss2021assemblingAccountability}
Moss, E.; Watkins, E.; Singh, R.; Elish, M.~C.; and Metcalf, J. 2021.
\newblock Assembling Accountability: Algorithmic Impact Assessment for the Public Interest.
\newblock \emph{SSRN Electronic Journal}.

\bibitem[{{National Institute of Standards and Technology}(2023{\natexlab{a}})}]{nist2023aiRisk}
{National Institute of Standards and Technology}. 2023{\natexlab{a}}.
\newblock {AI Risk Management Framework}.
\newblock Available at: www.nist.gov/itl/ai-risk-management-framework.

\bibitem[{{National Institute of Standards and Technology}(2023{\natexlab{b}})}]{equalAI2023nist}
{National Institute of Standards and Technology}. 2023{\natexlab{b}}.
\newblock {The EqualAI Algorithmic Impact Assessment Tool}.
\newblock Available at: www.equalai.org/aia.

\bibitem[{Nielsen and Landauer(1993)}]{nielsen1993}
Nielsen, J.; and Landauer, T.~K. 1993.
\newblock A Mathematical Model of the Finding of Usability Problems.
\newblock In \emph{Proceedings of the INTERACT and Conference on Human Factors in Computing Systems}, CHI '93, 206–213.

\bibitem[{Raji et~al.(2020)Raji, Smart, White, Mitchell, Gebru, Hutchinson, Smith-Loud, Theron, and Barnes}]{raji2020accountabilityGap}
Raji, I.~D.; Smart, A.; White, R.~N.; Mitchell, M.; Gebru, T.; Hutchinson, B.; Smith-Loud, J.; Theron, D.; and Barnes, P. 2020.
\newblock Closing the AI Accountability Gap: Defining an End-to-End Framework for Internal Algorithmic Auditing.
\newblock In \emph{Proceedings of the ACM Conference on Fairness, Accountability, and Transparency}, 33–44.

\bibitem[{Sadek et~al.(2024)Sadek, Constantinides, Quercia, and Mougenot}]{sadek2024guidelines}
Sadek, M.; Constantinides, M.; Quercia, D.; and Mougenot, C. 2024.
\newblock Guidelines for Integrating Value Sensitive Design in Responsible AI Toolkits.
\newblock In \emph{Proceedings of the ACM Conference on Human Factors in Computing Systems}, 1--20.

\bibitem[{Salda{\~n}a(2015)}]{saldana2015coding}
Salda{\~n}a, J. 2015.
\newblock \emph{{The Coding Manual for Qualitative Researchers}}.
\newblock Sage.

\bibitem[{Schiff et~al.(2020)Schiff, Rakova, Ayesh, Fanti, and Lennon}]{schiff2020principles}
Schiff, D.; Rakova, B.; Ayesh, A.; Fanti, A.; and Lennon, M. 2020.
\newblock Principles to Practices for Responsible AI: Closing the Gap.
\newblock arXiv:2006.04707.

\bibitem[{Selbst(2021)}]{selbst2021algorithmicImpact}
Selbst, A.~D. 2021.
\newblock An Institutional View of Algorithmic Impact.
\newblock \emph{Harvard Journal of Law \& Technology}, 35(1).

\bibitem[{Shelby et~al.(2023)Shelby, Rismani, Henne, Moon, Rostamzadeh, Nicholas, Yilla-Akbari, Gallegos, Smart, Garcia, and Virk}]{shelby2023sociotechnicalHarms}
Shelby, R.; Rismani, S.; Henne, K.; Moon, A.; Rostamzadeh, N.; Nicholas, P.; Yilla-Akbari, N.; Gallegos, J.; Smart, A.; Garcia, E.; and Virk, G. 2023.
\newblock Sociotechnical Harms of Algorithmic Systems: Scoping a Taxonomy for Harm Reduction.
\newblock In \emph{Proceedings of the AAAI/ACM Conference on AI, Ethics, and Society}, 723–741.

\bibitem[{Sherman and Eisenberg(2024)}]{sherman2023riskProfiles}
Sherman, E.; and Eisenberg, I. 2024.
\newblock AI Risk Profiles: A Standards Proposal for Pre-deployment AI Risk Disclosures.
\newblock In \emph{Proceedings of the AAAI Conference on Artificial Intelligence}, volume 38/21, 23047–23052.

\bibitem[{Skoric(2023)}]{skoric2023assessmentCriteria}
Skoric, V. 2023.
\newblock Critical Criteria for AI Impact Assessment: An Aggregated View.
\newblock \emph{SSRN Electronic Journal}.

\bibitem[{Sokol and Flach(2020)}]{sokol2020explanations}
Sokol, K.; and Flach, P. 2020.
\newblock One Explanation Does Not Fit All: The Promise of Interactive Explanations for Machine Learning Transparency.
\newblock \emph{KI - K\"{u}nstliche Intelligenz}, 34(2): 235–250.

\bibitem[{Stahl et~al.(2023)Stahl, Antoniou, Bhalla, Brooks, Jansen, Lindqvist, Kirichenko, Marchal, Rodrigues, Santiago, Warso, and Wright}]{stahl2023systematicReview}
Stahl, B.~C.; Antoniou, J.; Bhalla, N.; Brooks, L.; Jansen, P.; Lindqvist, B.; Kirichenko, A.; Marchal, S.; Rodrigues, R.; Santiago, N.; Warso, Z.; and Wright, D. 2023.
\newblock A Systematic Review of Artificial Intelligence Impact Assessments.
\newblock \emph{Artificial Intelligence Review}, 56(11): 12799–12831.

\bibitem[{{The Council of Europe}(2018)}]{algorithmsHumanRights_2018}
{The Council of Europe}. 2018.
\newblock {Algorithms and Human Rights. Study on the Human Rights Dimensions of Automated Data Processing Techniques and Possible Regulatory Implications}.
\newblock Available at: edoc.coe.int/en/internet/7589-algorithms-and-human-rights-study-on-the-human-rights-dimensions-of-automated-data-processing-techniques-and-possible-regulatory-implications.html.

\bibitem[{{The Danish Institute for Human Rights}(2023)}]{hria2023denmark}
{The Danish Institute for Human Rights}. 2023.
\newblock Introduction to Human Rights Impact Assessment.
\newblock Available at: www.humanrights.dk/tools/human-rights-impact-assessment-guidance-toolbox/introduction-human-rights-impact-assessment.

\bibitem[{{The Government of Canada}(2023)}]{admTool2023canada}
{The Government of Canada}. 2023.
\newblock Algorithmic Impact Assessment Tool.
\newblock Available at: open.canada.ca/aia-eia-js.

\bibitem[{{The High-Level Expert Group on Artificial Intelligence}(2020)}]{ALTAI_2020}
{The High-Level Expert Group on Artificial Intelligence}. 2020.
\newblock {The Assessment List for Trustworthy Artificial Intelligence}.
\newblock Available at: altai.insight-centre.org.

\bibitem[{{United Nations Educational, Scientific and Cultural Organization}(2023)}]{unesco2023ethicalImpactAssessment}
{United Nations Educational, Scientific and Cultural Organization}. 2023.
\newblock Ethical Impact Assessment. A Tool of the Recommendation on the Ethics of Artificial Intelligence.
\newblock Available at: www.unesco.org/en/articles/ethical-impact-assessment-tool-recommendation-ethics-artificial-intelligence.

\bibitem[{Vakkuri et~al.(2021)Vakkuri, Kemell, Jantunen, Halme, and Abrahamsson}]{eccola2021}
Vakkuri, V.; Kemell, K.-K.; Jantunen, M.; Halme, E.; and Abrahamsson, P. 2021.
\newblock ECCOLA — A Method for Implementing Ethically Aligned AI Systems.
\newblock \emph{Journal of Systems and Software}, 182: 111067.

\bibitem[{Wang et~al.(2023)Wang, Madaio, Kane, Kapania, Terry, and Wilcox}]{wang2023designing}
Wang, Q.; Madaio, M.; Kane, S.; Kapania, S.; Terry, M.; and Wilcox, L. 2023.
\newblock Designing Responsible AI: Adaptations of UX Practice to Meet Responsible AI Challenges.
\newblock In \emph{Proceedings of the ACM Conference on Human Factors in Computing Systems}, 1--16.

\bibitem[{Wang et~al.(2024)Wang, Kulkarni, Wilcox, Terry, and Madaio}]{farsight2024}
Wang, Z.~J.; Kulkarni, C.; Wilcox, L.; Terry, M.; and Madaio, M. 2024.
\newblock Farsight: Fostering Responsible AI Awareness During AI Application Prototyping.
\newblock In \emph{Proceedings of the ACM Conference on Human Factors in Computing Systems}, 1--40.

\bibitem[{Watkins et~al.(2021)Watkins, Moss, Metcalf, Singh, and Elish}]{Watkins2021}
Watkins, E.~A.; Moss, E.; Metcalf, J.; Singh, R.; and Elish, M.~C. 2021.
\newblock Governing Algorithmic Systems with Impact Assessments: Six Observations.
\newblock In \emph{Proceedings of the AAAI/ACM Conference on AI, Ethics, and Society}, 1010–1022.

\bibitem[{Weidinger et~al.(2023)Weidinger, Rauh, Marchal, Manzini, Hendricks, Mateos-Garcia, Bergman, Kay, Griffin, Bariach, Gabriel, Rieser, and Isaac}]{weidinger2023sociotechnical}
Weidinger, L.; Rauh, M.; Marchal, N.; Manzini, A.; Hendricks, L.~A.; Mateos-Garcia, J.; Bergman, S.; Kay, J.; Griffin, C.; Bariach, B.; Gabriel, I.; Rieser, V.; and Isaac, W. 2023.
\newblock Sociotechnical Safety Evaluation of Generative AI Systems.
\newblock arXiv:2310.11986.

\bibitem[{{World Economic Forum}(2022)}]{intendedUse2021}
{World Economic Forum}. 2022.
\newblock A Policy Framework for Responsible Limits on Facial Recognition. Use Case: Law Enforcement Investigations.
\newblock Available at: www.weforum.org/publications/a-policy-framework-for-responsible-limits-on-facial-recognition-use-case-law-enforcement-investigations-revised-2022.

\bibitem[{Wright and Wadhwa(2012)}]{wright2012privacyImpactAssessment}
Wright, D.; and Wadhwa, K. 2012.
\newblock Introducing a Privacy Impact Assessment Policy in the EU Member States.
\newblock \emph{International Data Privacy Law}, 3(1): 13--28.

\end{thebibliography}
\section{Appendix}

\subsection{(A) Summaries of the EU AI Act, NIST AI RMF, and ISO 42001 AI Risk Management Standard}
\label{subsec:summaries}

\subsubsection{EU AI Act.}
The Artificial Intelligence Act (AI Act) is a regulatory framework for artificial intelligence in European Union. It analyses AI systems in various uses and categorizes them based on the risk they present to users into unacceptable risk, high-risk, limited risk and minimal risk uses.

AI systems classified as high-risk and deployed by public bodies, private operators providing public services, or operators assessing creditworthiness and conducting risk assessments for life and health insurance must include reports with technical documentation and a mandatory assessment of their impact on fundamental rights. This impact assessment, conducted before deployment and updated as needed, should cover affected groups, risks of harm, human oversight, and risk management strategies. The information necessary for these reports is related to the following articles: 
\begin{itemize}
    \item Art. 9: Risk management system 
    \item Art. 10: Data and data governance
    \item Art. 11: Technical Documentation and 
    \item Art. 12: Record-keeping
    \item Art. 13: Transparency and provision of information to deployers
    \item Art. 14: Human oversight
    \item Art. 15: Accuracy, robustness and cybersecurity
    \item Art. 17: Quality management system
    \item Art. 16: Obligations of providers of high-risk AI systems
    \item Art. 18: Documentation keeping
    \item Art. 27: Fundamental rights impact assessment for high-risk AI systems
    \item Art. 50: Transparency obligations for providers and users of certain AI systems
    \item Art. 53: Obligations for providers of general-purpose AI models
    \item Art. 55: Obligations for providers of general-purpose AI models with systemic risk
    \item Art. 72: Post-market monitoring by providers and post-market monitoring plan for high-risk AI systems
\end{itemize}
Source: https://artificialintelligenceact.eu

\subsubsection{US NIST AI Risk Management Framework.}
The Artificial Intelligence Risk Management Framework \cite{nist2023aiRisk} is a voluntary guidance framework for organizations that design, develop, deploy, or use AI systems, aimed at helping them manage the risks associated with these systems. 
The framework is divided into two parts. The first part discusses how organizations can frame the risks related to AI and outlines the characteristics of trustworthy AI systems. These characteristics include being: valid and reliable, safe, secure and resilient, accountable and transparent, explainable and interpretable, privacy-enhanced, and fair. The second part, the core of the framework, describes four specific functions to help organizations address the risks of AI systems in practice: 
\begin{itemize}
    \item Govern – cultivate a culture of risk management
    \item Map – identify risks specific to context of use, 
    \item Measure – assess, analyze, and track identified risks, 
    \item Manage – prioritize risks and act upon their impact. 
\end{itemize}
Source: https://airc.nist.gov/home

\subsubsection{ISO/IEC 42001 AI Risk Management.}
This standard applies to organizations using, developing, monitoring, or providing AI products or services. Organizations shall establish a process to assess and document the potential consequences for individuals, groups of individuals, and societies that may result from the AI system through its life cycle. Specifically, the organization should assess whether an AI system affects the legal position or life opportunities of individuals, the physical or psychological well-being of individuals, universal human rights, and societies, and document:
\begin{itemize}
    \item the intended use of the reasonable foreseeable misuse of the AI system; 
    \item positive and negative impacts of the AI system to the relevant individuals and societies; 
    \item predictable failures, their potential impacts and measures taken to mitigate them; 
    \item relevant demographic groups the system is applicable to; 
    \item complexity of the system; 
    \item the role of humans in relationships with system, including human oversight capabilities, processes and tools, available to avoid negative impacts; 
    \item the resources of the AI system, including data, tools, systems, computing, and human resources (e.g., employment and staff skilling).
\end{itemize}

Source: https://www.iso.org/standard/81230.html

\subsection{(B) TEMPLATE ITERATIONS}
\begin{figure}[h!]
  \centering
  \includegraphics[width=0.5\textwidth]{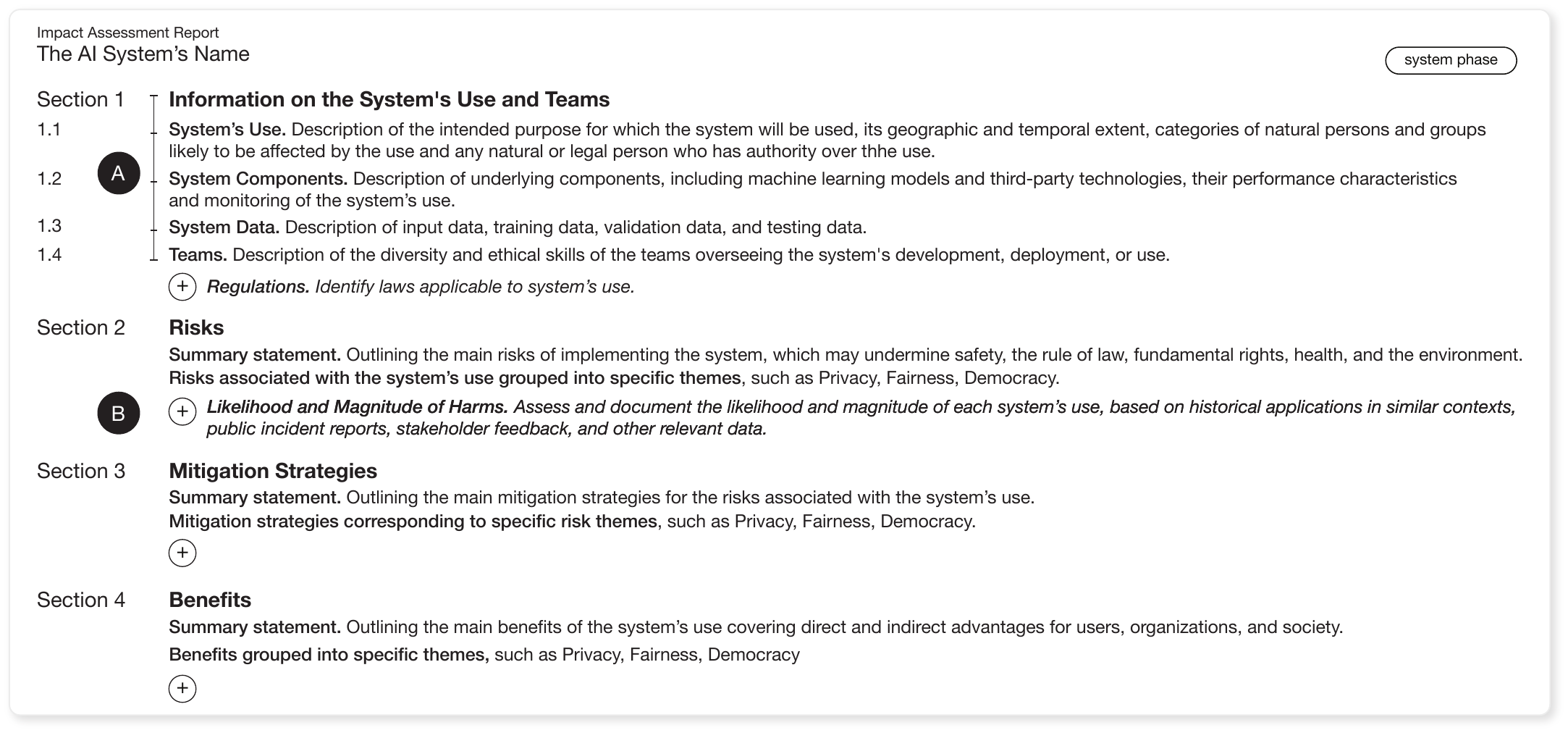}
  \caption{First version of the template for an Impact Assessment Report. Section 1 provides information on the system's use and teams; Section 2 lists potential risks; Section 3 lists mitigation strategies; and Section 4 outlines the anticipated benefits from the system's use.}
  \label{fig:template_v1}
\end{figure}

\noindent\textbf{V1} 
The primary concern with the first iteration with AI practitioners was the lack of evaluation details when the system is deployed in different contexts. P1 mentioned that \emph{`more details are needed about system's use in actual deployment and what are the risks''}. To resolve this issue, we added a new subsection ``System evaluation'', which details evaluation outcomes at each stage of the system's lifecycle: development, deployment, and use (Figure \ref{fig:template_v1}). 

\noindent\textbf{V2} 
The primary concerns with the second iteration were about the unclear description of the system's intended use and the limitations of this use. Participants stated that the template was either missing key components or its presentation made it hard for them to read. As P04 put it, \emph{``I'm missing description about the users of the system''}. 

First, to clarify the system's use, we divided its description into five subsections matching the risk assessment components identified by ~\citet{Golpayegani2023Risk}: purpose, capability, domain, user, and subject. Such a division provides all necessary information needed for conducting the risk assessment based on the EU AI Act. Users could specify these components using external dictionaries, such as the Vocabulary of AI Risks \cite{VAIR2023Framework}, which contains lists of descriptions for each of these concepts sourced directly from the Act. 

Second, after the "Evaluation" subsection, we added a sub-section to list system’s use limitations. This subsection can cover for example the reasonably foreseeable misuse: the use of an AI system in a way that is not in accordance with its intended use, but which may result from reasonably foreseeable human behaviour or interaction with other systems, including other AI systems.

\noindent\textbf{V3} 
The primary concerns with the third iteration were a lack of information about the governance of the use and a lack of a systematic way of reporting risks, mitigations, and benefits of this use based on the origins of the risks or benefits and whom they can impact. For example, P10, wanted \emph{``to see a bit more about the actual system to understand the system's transparency and its broader impact on society''}. Additionally, E04 stated that when she \emph{``tried to explain to our teams that we need to surface risks for organization, individuals, and groups of individuals, they found it too complicated and too remote to how currently we think about our business.''}.

To resolve the first issue, we added to the template the final section with risk reporting methods (e.g., dedicated email, the registered office) and compliance certifications. To resolve the second issue, we proposed to divide the ``Risks'', ``Mitigations'', 
and ``Benefits'' sections into three subsections each, aligning them with a three-layered framework for evaluating sociotechnical harms \cite{weidinger2023sociotechnical} for risks stemming from the \emph{system's capability} (i.e., the technical components of the AI system), \emph{human interaction} (i.e., experiences of people interacting with the AI system), and \emph{systemic impact} (i.e., societal, economic, and environmental impacts of the AI system's use). 

\noindent\textbf{V4} 
The primary concern with the fourth iteration was the lack of scaffolding elements supporting populating the template. To provide such elements for each section of the template, we examined guidelines and questions proposed in documentation standards \cite{selbst2021algorithmicImpact, gebru2021datasheets, holland2018dataset, bender2018data, mitchell2019model, sokol2020explanations, raji2020accountabilityGap}, questionnaires grounded in regulations \cite{ALTAI_2020, raiCrafting}, and checklists \cite{madaio2020co, Golpayegani2023Risk, skoric2023assessmentCriteria, equalAI2023nist} that structure ad-hoc practices in responsible AI. This resulted in a list of 32 statements that allow to systematically gather information about the system use, system components and data, the team involvement, and the risks, mitigations and benefits related to system's use \cite{reportProjectPage}.

\subsection{(C) LIST OF 32 STATEMENTS TO SYSTEMATICALLY COLLECT INFORMATION FOR THE IMPACT ASSESSMENT REPORT FROM STAKEHOLDERS}

Our project page \cite{reportProjectPage} outlines 32 statements to systematically collect information for the impact assessment report from stakeholders. Each guideline is phrased in simple, actionable language for easy understanding by both technical and non-technical stakeholders and is accompanied by a practical example.

\smallskip
\noindent\textbf{Statements 1--6 (System's Use).} To collect information about the \emph{system's use}, we considered five statements proposed by~\citet{Golpayegani2023Risk} and one by \citet{raiCrafting}. These statements are about the system's operational sector, users and subjects, thus providing a comprehensive description of the system's scope of use and compliance to date. Together, they are also useful for subsequent risk assessments of the use of the system ~\cite{Golpayegani2023Risk}. These statements are mapped to Section 1.1 of the final template (Figure~\ref{fig:final_template}).
\smallskip

\noindent\textbf{Statements 7--25 (System Components and Data).} These statements are about the \emph{system}: its components, its machine learning models, how to monitor it, and its data. We considered the statements proposed in the Model Card paper~\cite{mitchell2019model}, responsible AI guidelines~\cite{raiCrafting} and AI Fairness checklist \cite{madaio2020co}. This resulted in 19 statements covering various aspects, from the model's information and training data to its accuracy to its fairness. They are mapped to Sections 1.2 and 1.3 of the final template (Figure~\ref{fig:final_template})
\smallskip

\noindent\textbf{Statements 26 - 27 (Teams).} These statements concern the \emph{teams} behind the system use. We considered two statements from the Responsible AI guidelines~\cite{raiCrafting}, which are about team diversity and training on ethical values and regulations. They are mapped to Section 1.4 of Figure~\ref{fig:final_template}.
\smallskip

\noindent\textbf{Statements 28 - 30 (Risks and Mitigations).} These statements are about the potential \emph{risks} and harms caused by the system use. We considered three statements formulated by responsible AI guidelines~\cite{raiCrafting} and in the Model Card paper~\cite{mitchell2019model}. These statements cover various aspects such identification of use-related harms and procedures for their reporting. They are mapped to Sections 2 and 3 of the final template (Figure~\ref{fig:final_template}).
\smallskip

\noindent\textbf{Statements 31 - 32 (Benefits)}. These statements address the \emph{benefits} of the system's use for individuals, communities, organizations, and the planet. We sourced these statements from the NIST's AI RMF \cite{nist2023aiRisk, equalAI2023nist} and the ALTAI \cite{ALTAI_2020}. They are mapped to Section 4 of the final template (Figure~\ref{fig:final_template}).

\subsection{(D) POPULATED TEMPLATES}
\begin{figure*}[t!]
  \centering
  \includegraphics[width=0.96\textwidth]{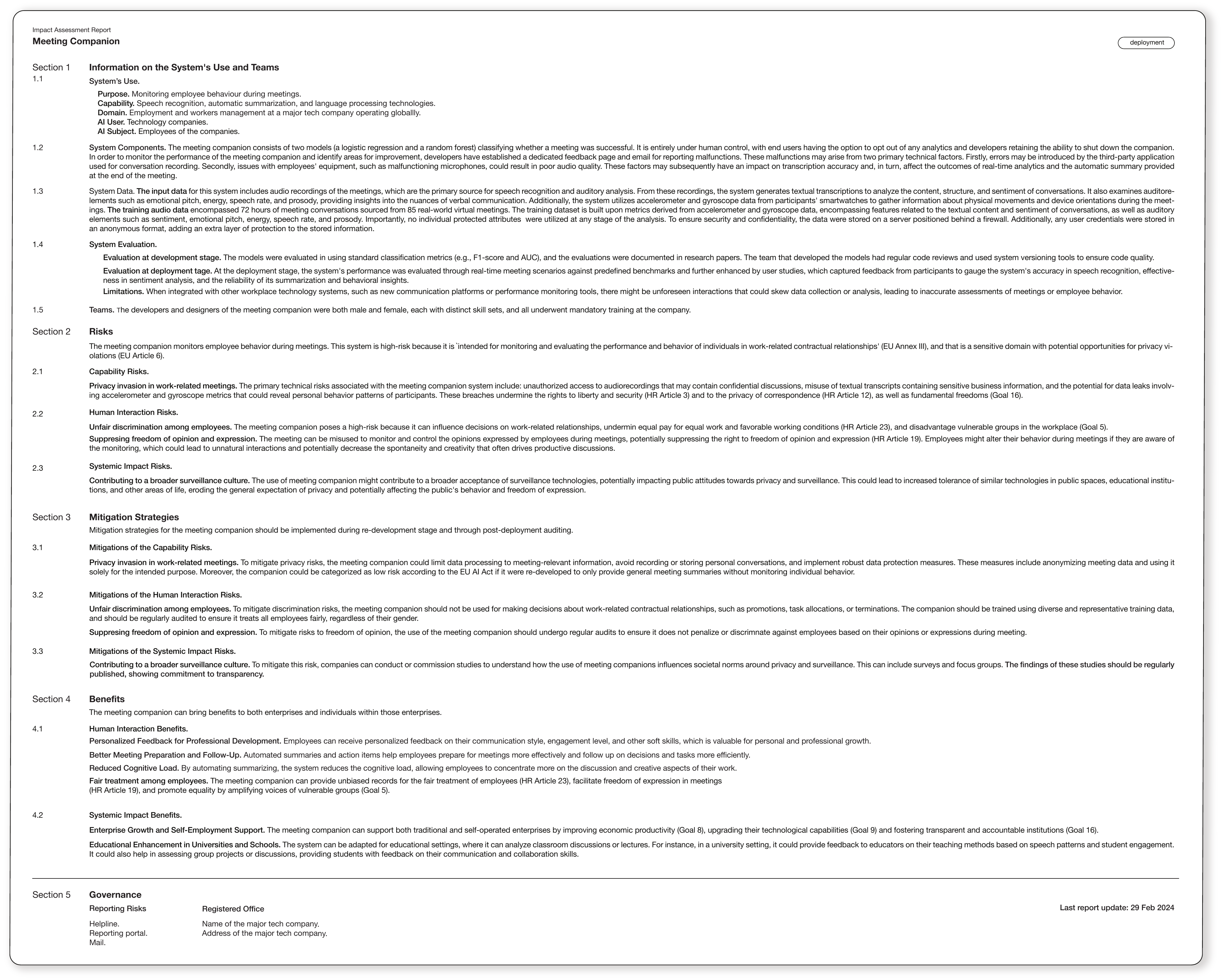}
  \caption{Impact assessment report for a meeting companion---an AI-based system aimed at monitoring employee behaviour during company meetings to improve meetings experience \cite{reportProjectPage}.}
  \label{fig:meeting_companion}
\end{figure*}

\begin{figure*}[t!]
  \centering
  \includegraphics[width=0.96\textwidth]{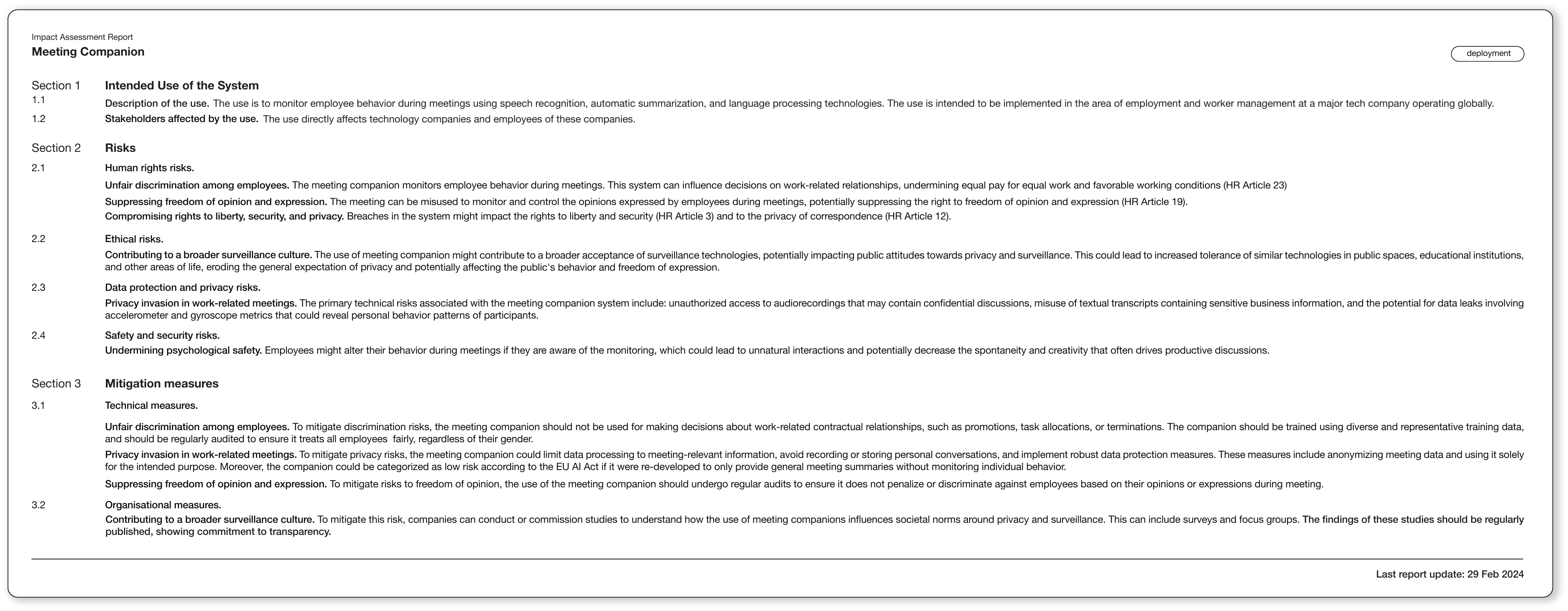}
  \caption{Impact assessment baseline for a meeting companion---an AI-based system aimed at monitoring employee behaviour during company meetings to improve meetings experience \cite{reportProjectPage}.}
  \label{fig:meeting_companion_baseline}
\end{figure*}

Figure \ref{fig:meeting_companion} and Figure \ref{fig:meeting_companion_baseline} present an impact assessment report for a meeting companion---an AI-based system designed to monitor employee behavior during company meetings to improve the meeting experience. The meeting companion is considered high-risk due to its capabilities in monitoring employee behavior during meetings (Figure \ref{fig:meeting_companion}, Section 1), which involves sensitive data collection that can infringe upon privacy rights and potentially affect public trust in technology within workplaces and society at large (Figure \ref{fig:meeting_companion}, Section 2; Figure \ref{fig:meeting_companion_baseline}, Section 2). This system also risks perpetuating inequalities by possibly impacting different racial groups unfavorably. However, these risks can be partially addressed by implementing measures such as utilizing anonymized data, employing diverse and representative historical datasets in training algorithms, and conducting regular audits to ensure the system's fairness and compliance (Figure \ref{fig:meeting_companion}, Section 3; Figure \ref{fig:meeting_companion_baseline}, Section 3). By adopting these mitigation strategies, the system can enhance organizational efficiency in conducting meetings and contribute to fairer treatment of employees (Figure \ref{fig:meeting_companion}, Section 4).

\subsection{(E) DEMOGRAPHICS SURVEY}
\label{app:demographics_survey}
\begin{itemize}
    \item How old are you? What is your gender?
    \item How many years of experience do you have in AI?
    \item What's your educational background?
    \item What is domain or sector of your work? (e.g., health)
    \item What is your current role? What kinds of AI systems do you work on (e.g., machine learning, computer vision)?
\end{itemize}

\end{document}